\documentclass[lettersize,journal]{IEEEtran}
\usepackage{array}
\newcolumntype{P}[1]{>{\centering\arraybackslash}m{#1}}
\usepackage[T1]{fontenc}
\usepackage[utf8]{inputenc} %
\usepackage{newtxtext,newtxmath} %
\usepackage{bm}  %
\usepackage{svg} 
\usepackage{amsmath,amsfonts}
\usepackage{algorithmic}
\usepackage{algorithm}
\usepackage{array}
\usepackage[caption=false,font=footnotesize,labelfont=sf,textfont=sf]{subfig}
\usepackage{textcomp}
\usepackage{stfloats}
\usepackage{url}
\usepackage{verbatim}
\usepackage{graphicx}
\usepackage{cite}
\usepackage{xcolor}
\usepackage[normalem]{ulem}

\usepackage{booktabs}
\usepackage[export]{adjustbox}
\usepackage[hidelinks]{hyperref}
\newcommand{\setdisplaymathspacing}{%
  \setlength{\abovedisplayskip}{0pt plus 0.5pt minus 0pt}%
  \setlength{\belowdisplayskip}{10pt plus 1pt minus 1pt}%
  \setlength{\abovedisplayshortskip}{0pt plus 0.5pt minus 0pt}%
  \setlength{\belowdisplayshortskip}{10pt plus 1pt minus 1pt}%
  \setlength{\jot}{2pt}%
}
\AtBeginDocument{\setdisplaymathspacing}
\AddToHook{cmd/normalsize/after}{\setdisplaymathspacing}
\AddToHook{cmd/small/after}{\setdisplaymathspacing}
\AddToHook{cmd/footnotesize/after}{\setdisplaymathspacing}
\AddToHook{env/equation/before}{\setdisplaymathspacing}
\makeatletter
\def\subsection{\@startsection{subsection}{2}{\z@}%
  {1.6ex plus 0.6ex minus 0.4ex}%
  {0.7ex plus .5ex minus 0ex}%
  {\normalfont\normalsize\itshape}}%
\makeatother
\ifdefined\pdfsuppresswarningpagegroup
\fi

\begin{document}

\title{Fourth-Order Cyclostationary Analysis of Power-Based Gardner-Type Timing Error Detectors in Coherent Optical Systems}

\author{Zhongxing Tian, Zeyu Feng, Huan Huang, Dongdong Zou,
Gordon Ning Liu, \textit{Member, IEEE}, \\ Gangxiang Shen, \textit{Senior Member, IEEE},
and Yi Cai*, \textit{Senior Member, IEEE}%
\thanks{This work was supported by the National Natural Science Foundation of China under Grant 62275185. (Corresponding author: Yi Cai.)}%
\thanks{Z. Tian, Z. Feng, Huan Huang, D. Zou, G. N. Liu, G. Shen, and Y. Cai are with Jiangsu Engineering Research Center of Novel Optical Fiber Technology and Communication Network, Suzhou Key Laboratory of Advanced Optical Communication Network Technology, and School of Electronic and Information Engineering, Soochow University, Suzhou, Jiangsu 215006, China (e-mail: zxtian@ieee.org; zyfeng1211@stu.suda.edu.cn; hhuang1799@gmail.com; ddzou@suda.edu.cn; gordonnliu@suda.edu.cn; shengx@suda.edu.cn; yicai@suda.edu.cn).}}

\maketitle
\setdisplaymathspacing

\begin{abstract}
Power-based nonlinear Gardner timing error detectors (TEDs) can enhance clock-tone (CT) extraction in low-roll-off and bandwidth-limited coherent optical systems. However, their nonlinear power-domain operations make the extracted CT components depend on higher-order cyclic statistics, which cannot be fully characterized by second-order cyclostationary analysis. In this paper, we develop a fourth-order cyclostationary analytical framework for power-based Gardner-type TEDs, using the square-Gardner TED (SG-TED) as a representative case. We show that the SG-TED CT originates from the symbol-rate cyclic component of the power-process autocorrelation function (CAF), revealing its fourth-order cyclic-statistical origin in the received complex field. Through moment-cumulant decomposition, the CT component is separated into a Wick-reducible term and a cumulant-related non-Gaussian term, which respectively explain its connection to the conventional Gardner/Godard mechanism and its modulation- and distribution-dependent behavior. The framework further characterizes the effects of pulse shaping, probabilistic shaping, polarization rotation, and polarization-mode dispersion (PMD), revealing CT-response characteristics fundamentally different from second-order TEDs. Numerical evaluations and waveform-level Monte Carlo simulations validate the analysis and demonstrate the framework as a unified statistical basis for SG-TED and related power-based Gardner-type TEDs.
\end{abstract}


\section{Introduction}

\IEEEPARstart{C}{lock} recovery (CR) is a key functional block in the digital signal processing (DSP) chain of high-speed coherent optical receivers \cite{ref1,ref2,ref3}. It estimates and compensates for the sampling clock mismatch between the digital-to-analog converter and the analog-to-digital converter, enabling the received waveform to be resampled at proper timing instants. Within a CR loop, the timing error detector (TED) extracts timing-error information from the received signal and provides the driving quantity for interpolation filtering \cite{ref4,ref5}. Therefore, the timing-information extraction capability of a TED is commonly characterized by the clock-tone (CT) strength at its output \cite{ref6,ref7,ref8}. A stronger CT generally indicates higher sensitivity to sampling-time deviations and a more reliable error signal for the CR loop.

Among existing CR schemes, the conventional Gardner \cite{ref9} and Godard \cite{ref10} algorithms are widely used because of their training-free operation, simple structure, and low implementation complexity \cite{ref11,ref12,ref13,ref14}. Their CT responses are closely related to the second-order cyclostationary property of the received signal \cite{ref15}, and are therefore sensitive to the pulse roll-off factor (ROF), bandwidth limitation, polarization rotation, and polarization-mode dispersion (PMD) \cite{ref13,ref16,ref17,ref18,ref19,ref20}. Under small ROFs or severe bandwidth constraints, the relevant symbol-rate cyclic spectral components are weakened, leading to CT attenuation at the TED output. Moreover, specific first-order PMD and polarization-coupling conditions, such as a differential group delay (DGD) of half a symbol period together with a 45$^\circ$ polarization rotation, can lead to complete CT fading \cite{ref18,ref19}. These phenomena can be systematically interpreted using the second-order cyclic autocorrelation function (CAF) and its frequency-domain counterpart, namely the spectral correlation function (SCF) \cite{ref21}.

The above two types of CT degradation have different physical origins and therefore motivate different treatments. The PMD- and polarization-induced singular fading originates from the destructive superposition of cyclic components carried by different polarization states, and can be mitigated by introducing a polarization correction module before the TED \cite{ref13,ref14,ref22,ref23}. In contrast, the CT weakening caused by low ROF or bandwidth limitation is rooted in the reduction of the second-order cyclic spectral overlap itself. To enhance timing-information extraction under such weak-CT conditions, several power-type modified Gardner TEDs have been proposed, including the square-Gardner TED (SG-TED) \cite{ref24}, interpolated-power Gardner (IP-Gardner) TED \cite{ref25}, and normalized interpolated-power Gardner (NIP-Gardner) TED \cite{ref8}. These methods introduce nonlinear transformations of the received waveform intensity and have shown improved CT extraction under low-ROF or bandwidth-limited conditions.

However, these nonlinear power-domain operations also change the statistical nature of the TED output. Taking the SG-TED as a representative example, it operates on the power process $|x(t)|^2$, and its output involves correlations of power samples. Therefore, the extracted CT component is no longer a second-order cyclic quantity of the complex field $x(t)$, but is governed by fourth-order cyclic statistics. As a result, the conventional second-order CAF/SCF framework \cite{ref21} is insufficient to fully characterize the CT generation mechanism of power-based Gardner-type TEDs, especially under different pulse shapes, modulation formats, symbol distributions, polarization rotations, and PMD conditions. A unified fourth-order cyclostationary framework is therefore needed.

Motivated by this gap, we develop a fourth-order cyclostationary analytical framework to analyze the generation mechanism and fading characteristics of the CT in power-based Gardner-type TEDs, using the SG-TED as a representative example. The framework reveals the fourth-order statistical origin of the SG-TED CT and enables analytical characterization under different pulse-shaping, modulation-format, signal-distribution, polarization-rotation, and PMD conditions. The derived results are validated by numerical evaluations and waveform-level simulations. The main contributions of this paper are summarized as follows:

\begin{itemize}
\setlength{\itemsep}{0pt}
\item[$\bullet$] We develop a fourth-order cyclostationary analytical framework for the SG-TED and show that its CT is generated by the symbol-rate cyclic component of the power-process CAF. We also review the conventional Gardner/Godard TEDs from a second-order cyclostationary perspective, providing a unified basis for comparing second-order and power-based Gardner-type timing mechanisms.

\item[$\bullet$] We decompose the SG-TED CT response into a Wick-reducible component and a cumulant-related non-Gaussian component. The former represents the fourth-order counterpart of the conventional Gardner/Godard timing mechanism, whereas the latter accounts for the modulation- and distribution-dependent response introduced by nonlinear power detection.

\item[$\bullet$] We systematically characterize the SG-TED CT response under pulse shaping, modulation format, probabilistic shaping, static polarization rotation, and first-order PMD. The results reveal CT-response characteristics fundamentally different from those of second-order TEDs and provide a statistical basis for understanding related power-based Gardner-type TEDs, such as IP-Gardner and NIP-Gardner.
\end{itemize}

The remainder of this paper is organized as follows. Section II introduces the signal model and the second-order cyclostationary analysis of Gardner/Godard TEDs. Section III develops the fourth-order cyclostationary analytical framework for the SG-TED. Section IV analyzes the effects of polarization rotation and first-order PMD. Section V presents numerical results and simulation validation. Finally, Section VI concludes the paper.

\section{Signal Model and Second-Order Cyclostationary Analysis of Gardner/Godard TEDs}

In this section, we establish the dual-polarization baseband signal model and introduce the second-order CAF and SCF. Based on these tools, we characterize the CT generation mechanism of Gardner/Godard TEDs and discuss their sensitivity to pulse roll-off and bandwidth limitation.

Considering the dual-polarization baseband signal in a coherent optical communication system, the transmitted waveforms on the two polarization tributaries are modeled as

\begin{equation}
\label{eq:word_1}
x(t)=\sum_k a_k g(t-kT_s-\tau),\quad
y(t)=\sum_k b_k g(t-kT_s-\tau).
\end{equation}
where $a_k$ and $b_k$ denote the complex modulation symbols carried by the two polarization tributaries, $T_s$ is the symbol period, $\tau$ represents the timing error, and $g(t)$ is the equivalent pulse response. The two symbol sequences are assumed to be mutually independent and identically distributed, satisfying

\begin{equation}
\label{eq:word_2}
\begin{gathered}
\mathbb{E}[a_k]=\mathbb{E}[b_k]=0,\quad
\mathbb{E}[a_k^2]=\mathbb{E}[b_k^2]=0,\\
\mathbb{E}[|a_k|^2]=\mathbb{E}[|b_k|^2]=\sigma_a^2.
\end{gathered}
\end{equation}
where $\mathbb{E}[\cdot]$ denotes mathematical expectation.

For a linearly modulated signal, its timing information can be characterized by second-order cyclostationary statistics. The CAF of $x(t)$ at cyclic frequency $\alpha$ is defined as

\begin{equation}
\label{eq:word_3}
R_x^\alpha(u)=\lim_{T_0\to\infty}\frac{1}{T_0}
\int_{-T_0/2}^{T_0/2}
\mathbb{E}\!\left[
x\!\left(t+\frac{u}{2}\right)
x^*\!\left(t-\frac{u}{2}\right)
\right]e^{-j2\pi\alpha t}\,dt.
\end{equation}
The corresponding SCF \cite{ref26,ref27} is given by

\begin{equation}
\label{eq:word_4}
\begin{aligned}
S_x^\alpha(f)&=\int_{-\infty}^{\infty}R_x^\alpha(u)e^{-j2\pi fu}\,du\\
&=\frac{\sigma_a^2}{T_s}
G\!\left(f+\frac{\alpha}{2}\right)
G^*\!\left(f-\frac{\alpha}{2}\right)e^{-j2\pi\alpha\tau},\quad
\alpha=\frac{m}{T_s}.
\end{aligned}
\end{equation}
where $G(f)$ is the Fourier transform of $g(t)$, and $m$ is an integer. Eq.~\eqref{eq:word_4} shows that the timing error $\tau$ introduces a phase rotation to the cyclic spectral component while leaving its magnitude unchanged. In particular, for the symbol-rate cyclic frequency $\alpha=1/T_s$, this component can be written as

\begin{equation}
\label{eq:word_6}
S_x^{1/T_s}(f;\tau)=S_x^{1/T_s}(f;\tau=0)e^{-j2\pi\tau/T_s}.
\end{equation}
Therefore, the symbol rate cyclic spectral component explicitly carries timing error information through its phase term.

\begin{figure}[!t]
\centering
\includegraphics[width=0.94\linewidth]{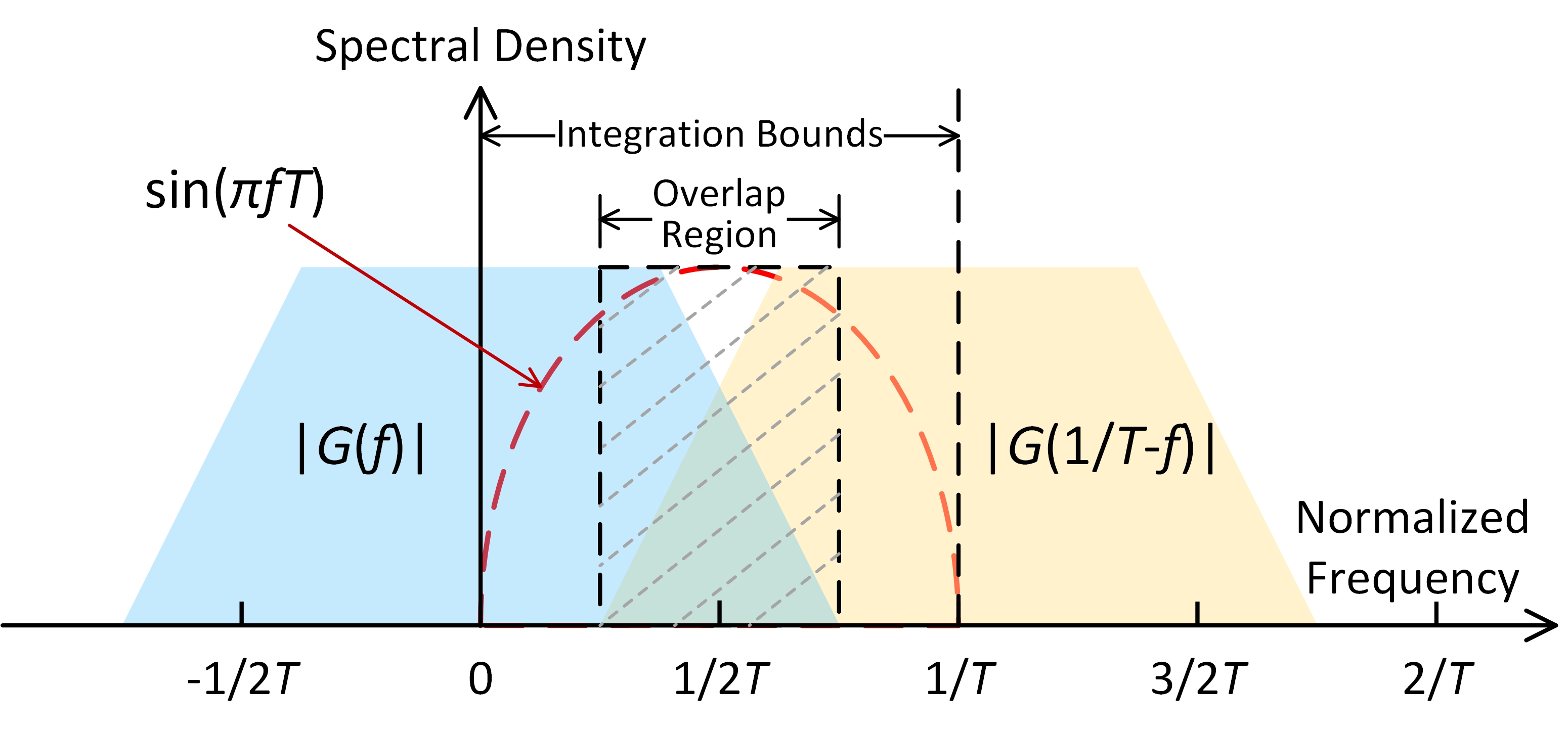}
\caption{Generation principle of the CT for Gardner and Godard TEDs}
\label{fig:word_1}
\end{figure}
The above analysis provides a direct interpretation of the Godard TED. Specifically, the Godard TED can be understood as extracting the symbol-rate cyclic component of the zero-lag CAF, i.e.,

\begin{equation}
\label{eq:word_7}
C_{\mathrm{Godard}}=R_x^{1/T_s}(0)=\int_{-\infty}^{\infty}S_x^{1/T_s}(f)\,df.
\end{equation}
When $g(t)$ is real-valued, the integral term at zero timing error is real-valued. Denoting this amplitude by $E_c$, the symbol-rate cyclic component and the corresponding Godard timing-error signal are given by

\begingroup
\small
\begin{equation}
\label{eq:word_8}
C_{\mathrm{Godard}}=E_c e^{-j2\pi\tau/T_s},\quad
\varepsilon_{\mathrm{Godard}}
=\operatorname{Im}\{C_{\mathrm{Godard}}\}
=-E_c\sin\!\left(\frac{2\pi\tau}{T_s}\right).
\end{equation}
\endgroup
Equivalently, a phase-based TED, commonly referred to as the BAJ detector \cite{ref28}, can be constructed by extracting the phase of the symbol-rate cyclic component. Under the adopted timing-error convention and within the principal timing range, it yields

\begin{equation}
\label{eq:word_10}
\varepsilon_{\mathrm{BAJ}}
=-\frac{1}{2\pi}\operatorname{unwrap}\{\arg(C_{\mathrm{Godard}})\}
=\frac{\tau}{T_s}.
\end{equation}

Unlike the Godard TED, the conventional Gardner TED employs an early-late structure in the time domain. Its TED can be expressed as \cite{ref9}

\begin{equation}
\label{eq:word_11}
\varepsilon_{\mathrm{Gardner}}
=\operatorname{Re}\!\left\{
x^*(t)\left[
x\!\left(t-\frac{T_s}{2}\right)
-x\!\left(t+\frac{T_s}{2}\right)
\right]\right\}.
\end{equation}
From the CAF/SCF perspective, the Gardner TED does not directly extract the zero-lag symbol-rate CAF component. Instead, it exploits an antisymmetric combination of the CAF values at half-symbol delays. According to the inverse relation between the CAF and SCF,

\begin{equation}
\label{eq:word_12}
R_x^\alpha(u)=\int_{-\infty}^{\infty}S_x^\alpha(f)e^{j2\pi fu}\,df,
\end{equation}
we have

\begin{equation}
\label{eq:word_13}
\begin{aligned}
R_x^\alpha\!\left(\frac{T_s}{2}\right)
-R_x^\alpha\!\left(-\frac{T_s}{2}\right)
&=\int_{-\infty}^{\infty}S_x^\alpha(f)
\left[e^{j\pi fT_s}-e^{-j\pi fT_s}\right]\,df\\
&=2j\int_{-\infty}^{\infty}S_x^\alpha(f)\sin(\pi fT_s)\,df.
\end{aligned}
\end{equation}
For the symbol-rate cyclic frequency $\alpha=1/T_s$, the dominant clock component extracted by the Gardner TED can therefore be expressed as

\begin{equation}
\label{eq:word_14}
C_{\mathrm{Gardner}}\propto
2j\int_{-\infty}^{\infty}S_x^{1/T_s}(f)\sin(\pi fT_s)\,df.
\end{equation}

Therefore, the Gardner TED extracts the same symbol-rate cyclostationary component as the Godard TED, but with an additional odd-symmetric frequency weighting factor $\sin(\pi fT_s)$ introduced by the early-late difference. As shown in Fig.~\ref{fig:word_1}, the Gardner CT strength is governed by the spectral overlap between the symbol-rate SCF and this weighting function. This explains its sensitivity to pulse ROF and bandwidth limitation: when the ROF is small or the signal spectrum is strongly bandwidth-limited, the effective overlap region is reduced, leading to CT attenuation.

\section{Fourth-Order Cyclostationary Analysis of the Square-Gardner TED}

In this section, we develop a fourth-order cyclostationary analytical framework for the SG-TED by relating its output to the CAF of the received power process. This formulation reveals the statistical origin of CT generation and decomposes the relevant correlation into Wick-reducible and cumulant-related components, thereby clarifying the underlying timing mechanism.

\subsection{SG-TED and Fourth-Order CAF}

Applying the Gardner early--late operation to the received power process $p(t)=|x(t)|^2$, the continuous-time SG-TED output, with $u_0=T_s/2$, is given by

\begin{equation}
\label{eq:word_16}
\varepsilon_{\mathrm{SG}}=p(t)\left[p(t-u_0)-p(t+u_0)\right].
\end{equation}
The SG-TED output contains products of two power samples. Therefore, its ensemble-averaged response involves fourth-order moments of the original complex field, i.e.,

\begin{equation}
\label{eq:word_17}
\mathbb{E}\!\left[|x(t_1)|^2|x(t_2)|^2\right]
=\mathbb{E}\!\left[x(t_1)x^*(t_1)x(t_2)x^*(t_2)\right].
\end{equation}
To characterize this quantity, we define the CAF of the power process as

\begin{equation}
\label{eq:word_18}
\begin{aligned}
R_p^\alpha(u)
&=\lim_{T_0\to\infty}\frac{1}{T_0}
\int_{-T_0/2}^{T_0/2}
\mathbb{E}\!\left[
p\!\left(t+\frac{u}{2}\right)
p^*\!\left(t-\frac{u}{2}\right)
\right]e^{-j2\pi\alpha t}\,dt\\
&=\lim_{T_0\to\infty}\frac{1}{T_0}
\int_{-T_0/2}^{T_0/2}
\mathbb{E}\!\left[
x_+x_+^*x_-x_-^*
\right]e^{-j2\pi\alpha t}\,dt.
\end{aligned}
\end{equation}
where $x_+=x(t+u/2)$ and $x_-=x(t-u/2)$.

Using the time-shift property of Fourier coefficients, the cyclic component of the SG-TED output at cyclic frequency $\alpha$ can be written as

\begin{equation}
\label{eq:word_19}
C_{\mathrm{SG}}^\alpha(u_0)
=-2j\sin(\pi\alpha u_0)R_p^\alpha(u_0).
\end{equation}
The derivation of Eq.~\eqref{eq:word_19} is given in Appendix~\ref{app:sg_cyclic}. 
For $\alpha=1/T_s$ and $u_0=T_s/2$,

\begin{equation}
\label{eq:word_20}
C_{\mathrm{SG}}=-2jR_p^{1/T_s}\!\left(\frac{T_s}{2}\right).
\end{equation}
The sign depends on the adopted early-late convention. Eq.~\eqref{eq:word_20} shows that the SG-TED clock component is directly determined by the symbol-rate cyclic component of the power-process CAF at the half-symbol lag.

\subsection{Fourth-Order Decomposition and Clock-Tone Generation Mechanism}

We next decompose the fourth-order moment in Eq.~\eqref{eq:word_18}. For the linearly modulated signal, define

\begin{equation}
\label{eq:word_21}
g_{k,+}=g(t+u/2-kT_s-\tau),\quad
g_{k,-}=g(t-u/2-kT_s-\tau).
\end{equation}
Then

\begin{equation}
\label{eq:word_22}
x_+=\sum_k a_k g_{k,+},\qquad
x_-=\sum_k a_k g_{k,-}.
\end{equation}
For independent and identically distributed zero-mean proper complex symbols, the fourth-order cumulant is defined as

\begin{equation}
\label{eq:word_23}
\kappa_a
=
\mathbb{E}\!\left[|a_k|^4\right]
-
2\left(\mathbb{E}\!\left[|a_k|^2\right]\right)^2.
\end{equation}
where $\sigma_a^2=\mathbb{E}[|a_k|^2]$ denotes the symbol variance. Applying the moment-cumulant expansion yields

\begin{equation}
\label{eq:word_24}
\begin{aligned}
\mathbb{E}\!\left[|x_+|^2|x_-|^2\right]
&=m\!\left(t+\frac{u}{2}\right)m\!\left(t-\frac{u}{2}\right)
+\left|w_x(t,u)\right|^2+\kappa_a d(t,u).
\end{aligned}
\end{equation}
where

\begin{equation}
\label{eq:word_25}
m(t)=\sigma_a^2\sum_k |g(t-kT_s-\tau)|^2.
\end{equation}
is the instantaneous average power envelope,

\begin{equation}
\label{eq:word_26}
w_x(t,u)=\sigma_a^2\sum_k
g(t+u/2-kT_s-\tau)g^*(t-u/2-kT_s-\tau).
\end{equation}
is the second-order field-correlation kernel, and

\begin{equation}
\label{eq:word_27}
d(t,u)=\sum_k
|g(t+u/2-kT_s-\tau)|^2
|g(t-u/2-kT_s-\tau)|^2.
\end{equation}
is the fourth-order pulse-overlap kernel. The detailed derivation of Eq.~\eqref{eq:word_24} is given in Appendix~\ref{app:fourth_moment}.

Eq.~\eqref{eq:word_24} separates the fourth-order power correlation into two Wick-reducible contributions and one cumulant contribution. The first term is the product of average power envelopes. The second term, $|w_x(t,u)|^2$, is a Wick-reducible contribution because it is fully determined by the second-order field-correlation kernel $w_x(t,u)$. The kernel is the time-periodic second-order field-correlation kernel of the received complex waveform. Its symbol-rate Fourier coefficients constitute the CAF components exploited by the Gardner/Godard TEDs. In contrast, the SG-TED does not use $w_x(t,u)$ directly, its Wick-reducible contribution is given by the symbol-rate Fourier coefficient of $|w_x(t,T_s/2)|^2$. Therefore, the SG-TED Wick term is generated from the same second-order field-correlation kernel, but through a fourth-order power-correlation operation. The last term, $\kappa_a d(t,u)$, depends on the fourth-order cumulant of the modulation symbols and is therefore absent for a proper complex Gaussian signal.

For a $T_s$-periodic function $F(t,u)$ with respect to the center time $t$, its Fourier coefficient at the symbol-rate cyclic frequency is defined as

\begin{equation}
\label{eq:word_28}
F^{1/T_s}(u)=\frac{1}{T_s}\int_{-T_s/2}^{T_s/2}
F(t,u)e^{-j2\pi t/T_s}\,dt.
\end{equation}
Using Eq.~\eqref{eq:word_24}, the symbol-rate cyclic component of the power-process CAF can be expressed as

\begin{equation}
\label{eq:word_29}
R_p^{1/T_s}(u)=M^{1/T_s}(u)+W^{1/T_s}(u)+\kappa_aD^{1/T_s}(u).
\end{equation}
where

\begin{equation}
\label{eq:word_30}
\begin{aligned}
M^{1/T_s}(u)&=\left[m(t+u/2)m(t-u/2)\right]_{1/T_s},\\
W^{1/T_s}(u)&=\left[\left|w_x(t,u)\right|^2\right]_{1/T_s},
\end{aligned}
\end{equation}
and

\begin{equation}
\label{eq:word_31}
D^{1/T_s}(u)=\left[d(t,u)\right]_{1/T_s}.
\end{equation}
Here, $[\cdot]_{1/T_s}$ denotes the symbol rate Fourier coefficient defined in Eq.~\eqref{eq:word_28}. For the SG-TED, $u=u_0=T_s/2$. In this case, the average-power product has a period of $T_s/2$, and hence contains no symbol-rate Fourier component, i.e.,

\begin{equation}
\label{eq:word_32}
M^{1/T_s}\!\left(T_s/2\right)=0.
\end{equation}
The proof is given in Appendix~\ref{app:vanishing_average}. Therefore,

\begin{equation}
\label{eq:word_33}
R_p^{1/T_s}\!\left(\frac{T_s}{2}\right)
=W^{1/T_s}\!\left(\frac{T_s}{2}\right)
+\kappa_aD^{1/T_s}\!\left(\frac{T_s}{2}\right).
\end{equation}
Combining Eqs.~\eqref{eq:word_20} and \eqref{eq:word_33}, the symbol-rate clock component of the SG-TED becomes

\begin{equation}
\label{eq:word_34}
C_{\mathrm{SG}}=-2j\left[
W^{1/T_s}\!\left(\frac{T_s}{2}\right)
+\kappa_aD^{1/T_s}\!\left(\frac{T_s}{2}\right)
\right].
\end{equation}
When the timing error $\tau$ is explicitly considered, the temporal shift in $x(t)$ introduces the phase factor $e^{-j2\pi\tau/T_s}$. Thus

\begin{equation}
\label{eq:word_35}
\begin{aligned}
C_{\mathrm{SG}}(\tau)
&=-2j\left[
W_0^{1/T_s}\!\left(\frac{T_s}{2}\right)
+\kappa_aD_0^{1/T_s}\!\left(\frac{T_s}{2}\right)
\right]e^{-j2\pi\tau/T_s}.
\end{aligned}
\end{equation}
where $W_{0}^{1/T_{s}}$ and $D_{0}^{1/T_{s}}$ denote the corresponding coefficients at $\tau=0$. The CT strength is then

\begin{equation}
\label{eq:word_36}
\begin{aligned}
G_{\mathrm{SG}}
&=4\left|
W_0^{1/T_s}\!\left(\frac{T_s}{2}\right)
+\kappa_aD_0^{1/T_s}\!\left(\frac{T_s}{2}\right)
\right|^2=4|C_0+C_\kappa|^2.
\end{aligned}
\end{equation}

Eqs.~\eqref{eq:word_34}--\eqref{eq:word_36} give the fourth-order cyclostationary interpretation of the SG-TED. The Wick-reducible component $W^{1/T_s}(T_s/2)$ is linked to the second-order field-correlation kernel. The cumulant component $\kappa_aD^{1/T_s}(T_s/2)$, in contrast, is governed by the non-Gaussian fourth-order statistics of the transmitted symbols and the fourth-order pulse-overlap kernel.

\subsection{Pulse-Shape and Modulation Dependence of the CT Response}

The fourth-order representation in Eq.~\eqref{eq:word_36} separates the effects
of pulse shaping and symbol statistics on the SG-TED CT response.
Specifically, the coefficients
$W_0^{1/T_s}(T_s/2)$ and $D_0^{1/T_s}(T_s/2)$ are governed by the pulse
waveform, whereas $\kappa_a$ is determined by the symbol distribution.
The CT response therefore depends jointly on the pulse shape and the
fourth-order statistics of the modulation symbols.

The pulse-shape dependence is embodied in the kernels $w_x(t,u)$ and
$d(t,u)$. Pulse-shaping parameters, including the roll-off factor,
pulse width, and bandwidth constraints, modify the symbol-rate Fourier
coefficients of $|w_x(t,T_s/2)|^2$ and $d(t,T_s/2)$, and hence the CT
strength. This mechanism differs from that of conventional
Gardner/Godard TEDs, whose CT response is governed by a second-order
cyclic spectral-overlap relation. In the SG-TED, the nonlinear power
operation reshapes the relevant cyclostationary structure and gives
rise to fourth-order pulse-overlap terms.

The modulation dependence enters through the fourth-order cumulant defined in
Eq.~\eqref{eq:word_23}. For normalized constellations satisfying
$\mathbb{E}[|a_k|^2]=1$, the unit average symbol energy can be directly
substituted into that definition. Consequently,
modulation formats and symbol distributions with the
same average symbol energy may yield different SG-TED CT strengths.
For a proper complex Gaussian signal, $\kappa_a=0$, and the cumulant
contribution vanishes. By contrast, $\kappa_a$ is generally nonzero
for finite-alphabet constellations. Probabilistic shaping further
changes $\mathbb{E}[|a_k|^4]$ and thereby modifies the relative weight
of the cumulant contribution.

These results indicate that the SG-TED should be interpreted through the
symbol-rate cyclic component of the fourth-order power correlation, rather
than as a simple power-domain counterpart of the conventional Gardner/Godard
TEDs. The Wick-reducible term retains the connection to the classical
second-order timing mechanism, whereas the cumulant term introduces the
modulation-dependent contribution, thereby explaining the CT dependence on
pulse shape, bandwidth, and symbol distribution observed in the following
numerical results.

\section{Effects of Polarization Rotation and PMD on Gardner/Godard and SG-TEDs}

In this section, we analyze the CT responses of these TEDs under static polarization rotation and first-order PMD model. The comparison highlights the different polarization-dependent fading mechanisms of second-order and fourth-order timing extraction.

\subsection{Static Polarization Rotation}

Considering the signal entering a single-polarization TED branch after a static polarization rotation,

\begin{equation}
\label{eq:word_40}
z_{\theta,\phi}(t)=\cos\theta\,x(t)+e^{j\phi}\sin\theta\,y(t),
\end{equation}
where $\theta$ is the azimuth, and $\phi$ is the ellipticity. Let $c=\cos\theta$ and $s=\sin\theta$. Since the two polarization signals are assumed to be independent and identically distributed, the second-order SCF of $z_{\theta,\phi}(t)$ is

\begin{equation}
\label{eq:word_41}
S_z^\alpha(f)=c^2S_x^\alpha(f)+s^2S_x^\alpha(f)=S_x^\alpha(f).
\end{equation}
Thus, static polarization rotation does not change the second-order cyclic spectral component. Consequently, the CT strengths of Gardner/Godard TEDs remain unchanged under static polarization rotation, i.e.,

\begin{equation}
\label{eq:word_42}
\begin{aligned}
G_{\mathrm{Godard}}(\theta,\phi)
&=G_{\mathrm{Godard}}(0,0),\\
G_{\mathrm{Gardner}}(\theta,\phi)
&=G_{\mathrm{Gardner}}(0,0).
\end{aligned}
\end{equation}
This invariance follows from the fact that Gardner/Godard TEDs extract timing information from second-order cyclic statistics, which are preserved under a unitary mixing of two statistically identical and independent polarization tributaries.

The SG-TED has a different dependence because its CT response contains a fourth-order cumulant contribution. Under Eq.~\eqref{eq:word_40}, the effective symbol carried by the rotated signal is

\begin{equation}
\label{eq:word_43}
\tilde{a}_k=ca_k+e^{j\phi}sb_k.
\end{equation}
Although its second-order power is unchanged,

\begin{equation}
\label{eq:word_44}
\begin{aligned}
\mathbb{E}\!\left[|\tilde{a}_k|^2\right]
&=c^2\mathbb{E}\!\left[|a_k|^2\right]
+s^2\mathbb{E}\!\left[|b_k|^2\right]\\
&\quad+cse^{j\phi}\mathbb{E}\!\left[a_kb_k^*\right]
+cse^{-j\phi}\mathbb{E}\!\left[b_ka_k^*\right]
=\sigma_a^2,
\end{aligned}
\end{equation}
its fourth-order cumulant becomes

\begin{equation}
\label{eq:word_45}
\kappa_{\tilde{a}}=|c|^4\kappa_a+|se^{-j\phi}|^4\kappa_b
=\left(1-\frac{1}{2}\sin^2(2\theta)\right)\kappa_a.
\end{equation}
Therefore, using the SG-TED CT expression derived in Section III, the CT strength under static polarization rotation can be written as

\begin{equation}
\label{eq:word_46}
\begin{aligned}
G_{\mathrm{SG}}(\theta)
&=4\left|
W_0^{1/T_s}\!\left(\frac{T_s}{2}\right)
+\left(1-\frac{1}{2}\sin^2(2\theta)\right)
\kappa_aD_0^{1/T_s}\!\left(\frac{T_s}{2}\right)
\right|^2\\
&=4|C_0+\eta(\theta)C_\kappa|^2.
\end{aligned}
\end{equation}
where

\begin{equation}
\label{eq:word_47}
\eta(\theta)=\cos^4\theta+\sin^4\theta
=1-\frac{1}{2}\sin^2(2\theta).
\end{equation}
This result shows that static polarization rotation does not affect the Wick-reducible component, but scales the cumulant-related component. Hence, the SG-TED is generally not invariant to static polarization rotation when the cumulant term is significant.

\subsection{First-Order PMD Response}

We next analyze the effect of first-order PMD \cite{ref29}. In the principal-state representation, the PMD channel can be written as

\begin{equation}
\label{eq:word_48}
\mathbf{H}(f)=\mathbf{U}(\theta,\phi)
\begin{bmatrix}
e^{-j\pi f\Delta\tau} & 0\\
0 & e^{j\pi f\Delta\tau}
\end{bmatrix}
\mathbf{U}^{H}(\theta,\phi).
\end{equation}
where $\Delta\tau$ is the DGD and

\begin{equation}
\label{eq:word_49}
\mathbf{U}(\theta,\phi)=
\begin{bmatrix}
c & -e^{j\phi}s\\
e^{-j\phi}s & c
\end{bmatrix}.
\end{equation}

For the Gardner/Godard TEDs, the relevant quantity is the second-order symbol rate cyclic spectrum. Assuming independent and identically distributed input polarization signals, the input SCF matrix is

\begin{equation}
\label{eq:word_50}
\mathbf{S}_{\mathrm{in}}^\alpha(f)=S_x^\alpha(f)\mathbf{I}.
\end{equation}
After the PMD channel, the output SCF matrix is

\begin{equation}
\label{eq:word_51}
\mathbf{S}_{\mathrm{out}}^\alpha(f)
=\mathbf{H}\!\left(f+\frac{\alpha}{2}\right)
\mathbf{S}_{\mathrm{in}}^\alpha(f)
\mathbf{H}^{H}\!\left(f-\frac{\alpha}{2}\right).
\end{equation}
Substituting Eq.~\eqref{eq:word_50} into Eq.~\eqref{eq:word_51}, the cyclic spectrum observed at the first output branch can be written as

\begin{equation}
\label{eq:word_52}
S_{x,\mathrm{PMD}}^\alpha(f)
=\eta_2(\alpha,\theta,\Delta\tau)S_x^\alpha(f),
\end{equation}
where

\begin{equation}
\label{eq:word_53}
\eta_2(\alpha,\theta,\Delta\tau)
=c^2e^{-j\pi\alpha\Delta\tau}+s^2e^{j\pi\alpha\Delta\tau}.
\end{equation}
The phase $\phi$ does not appear in Eq.~\eqref{eq:word_53}, because it is canceled in the diagonal element of the second-order SCF transformation under the assumed identical and independent tributaries. The derivation of Eqs.~\eqref{eq:word_48}--\eqref{eq:word_53} is given in Appendix~\ref{app:pmd_second_order}. For $\alpha=1/T_s$, the magnitude of the PMD-induced second-order fading factor is

\begin{equation}
\label{eq:word_54}
\Gamma(\theta,\Delta\tau)
=\left|\eta_2(1/T_s,\theta,\Delta\tau)\right|^2
=1-\sin^2(2\theta)\sin^2(\pi\Delta\tau/T_s).
\end{equation}
Therefore, the CT strengths of the Godard and Gardner TEDs under first-order PMD are

\begin{equation}
\label{eq:word_55}
\begin{aligned}
G_{\mathrm{Godard}}^{\mathrm{PMD}}
&=\Gamma(\theta,\Delta\tau)G_{\mathrm{Godard}}^0,\\
G_{\mathrm{Gardner}}^{\mathrm{PMD}}
&=\Gamma(\theta,\Delta\tau)G_{\mathrm{Gardner}}^0.
\end{aligned}
\end{equation}
where $G_{\mathrm{Godard}}^0$ and $G_{\mathrm{Gardner}}^0$ denote the corresponding CT strengths without PMD. Eqs.~\eqref{eq:word_53}--\eqref{eq:word_55} indicate that complete second-order CT fading occurs when

\begin{equation}
\label{eq:word_56}
\theta=(2m+1)\frac{\pi}{4},\qquad
\Delta\tau=(2\ell+1)\frac{T_s}{2},\qquad m,\ell\in\mathbb{Z}.
\end{equation}
Under these conditions, the two principal-state contributions have equal amplitudes and opposite symbol-rate cyclic phases, leading to complete cancellation of the second-order CT.

The SG-TED requires a different treatment because it operates on the power process. From the first row of the PMD matrix in Eq.~\eqref{eq:word_48}, the observed field can be expressed in the time domain as

\begin{equation}
\label{eq:word_57}
x_{\mathrm{PMD}}(t)=\sum_k a_k h_x(t-kT_s-\tau)
+\sum_k b_k h_y(t-kT_s-\tau),
\end{equation}
where

\begin{equation}
\label{eq:word_58}
\begin{aligned}
h_x(t)&=c^2g\!\left(t-\frac{\Delta\tau}{2}\right)
+s^2g\!\left(t+\frac{\Delta\tau}{2}\right),\\
h_y(t)&=cse^{j\phi}
\left[g\!\left(t-\frac{\Delta\tau}{2}\right)
-g\!\left(t+\frac{\Delta\tau}{2}\right)\right].
\end{aligned}
\end{equation}
Thus, $\phi$ only appears as a common phase factor in $h_y(t)$. Since the SG-TED CT strength depends on products such as $h_{y,k,+}h_{y,k,-}^{*}$ and $\left|h_{y,k,\pm }\right|^{2}$, this common phase factor cancels in the fourth-order decomposition. Hence, under the assumed proper-complex and independent tributaries, $\phi$ does not affect the resulting CT strength.

Applying the fourth-order decomposition in Section III to $x_{\mathrm{PMD}}(t)$, the symbol-rate power-process CAF at the SG-TED lag $u_0=T_s/2$ can be written as

\begin{equation}
\label{eq:word_59}
R_{p,\mathrm{PMD}}^{1/T_s}\!\left(\frac{T_s}{2}\right)
=W_{\mathrm{PMD}}^{1/T_s}\!\left(\frac{T_s}{2}\right)
+\kappa_aD_{\mathrm{PMD}}^{1/T_s}\!\left(\frac{T_s}{2}\right).
\end{equation}
Here, $W_{\mathrm{PMD}}^{1/T_s}$ is the Wick-reducible contribution, and $D_{\mathrm{PMD}}^{1/T_s}$ is the cumulant-related fourth-order pulse-overlap contribution. Both are computed from the PMD-distorted pulse responses $h_x(t)$ and $h_y(t)$. Their explicit forms are derived in Appendix~\ref{app:pmd_fourth_order}.

The corresponding SG-TED CT strength is

\begin{equation}
\label{eq:word_60}
\begin{aligned}
G_{\mathrm{SG}}^{\mathrm{PMD}}
&=4\left|
W_{\mathrm{PMD}}^{1/T_s}\!\left(\frac{T_s}{2}\right)
+\kappa_aD_{\mathrm{PMD}}^{1/T_s}\!\left(\frac{T_s}{2}\right)
\right|^2=4|C_{\mathrm{PMD}}+C_{\kappa,\mathrm{PMD}}|^2.
\end{aligned}
\end{equation}
Eq.~\eqref{eq:word_60} shows that the SG-TED PMD response cannot be represented by the second-order fading factor $\Gamma(\theta,\Delta\tau)$ alone. Instead, it is determined by the PMD-distorted effective pulse responses, the resulting fourth-order cyclic coefficients, and the modulation cumulant $\kappa_a$. This explains the reason that the SG-TED can exhibit a PMD-dependent fading behavior different from that of the Gardner/Godard TED.

\section{Numerical Results and Simulation Validation}

In this section, the analytical results derived in the preceding sections are numerically evaluated, with Monte-Carlo simulations conducted for waveform-level validation. We first compare the CT strengths of Gardner TED and the SG-TED under RRC and Gaussian pulse-shaping conditions. The effects of modulation format, symbol distribution, static polarization rotation, and first-order PMD are then examined to clarify the roles of the Wick-reducible and cumulant-related components and to reveal the distinct polarization-dependent characteristics of the SG-TED. Finally, IP-Gardner and NIP-Gardner schemes are briefly discussed.

\subsection{Simulation Setup}

The analytical curves are obtained by directly evaluating the derived CT expressions in a normalized symbol-time domain and therefore depend only on dimensionless pulse-shaping and statistical parameters. Monte-Carlo simulations are performed with a representative coherent optical system configuration to generate finite-length discrete-time waveforms and validate the analytical predictions under practical impairment-limited conditions. The main simulation parameters are summarized in Table~\ref{tab:word_1}. Unless otherwise specified, only the pulse-shaping parameter, modulation format, polarization-rotation angle, or DGD is varied for each specific case.

\begin{table}[!t]
\caption{Parameters used for simulation validation.}
\label{tab:word_1}
\centering
\begin{tabular}{ll}
\hline
Parameter & Value \\
\hline
Baud rate & 60 GBd \\
Sampling rate & 120 GSa/s \\
Number of symbols & $2^{16}$ \\
OSNR & 30 dB \\
Laser linewidth & 200 kHz (Tx/Rx 100 kHz) \\
Frequency offset & 1 GHz \\
Modulation & QAM/PS-QAM \\
Pulse shaping & RRC/Gaussian \\
\hline
\end{tabular}
\end{table}

\subsection{Effects of Pulse-Shaping Parameters}

\begin{figure}[!t]
\centering
\includegraphics[width=0.9\linewidth]{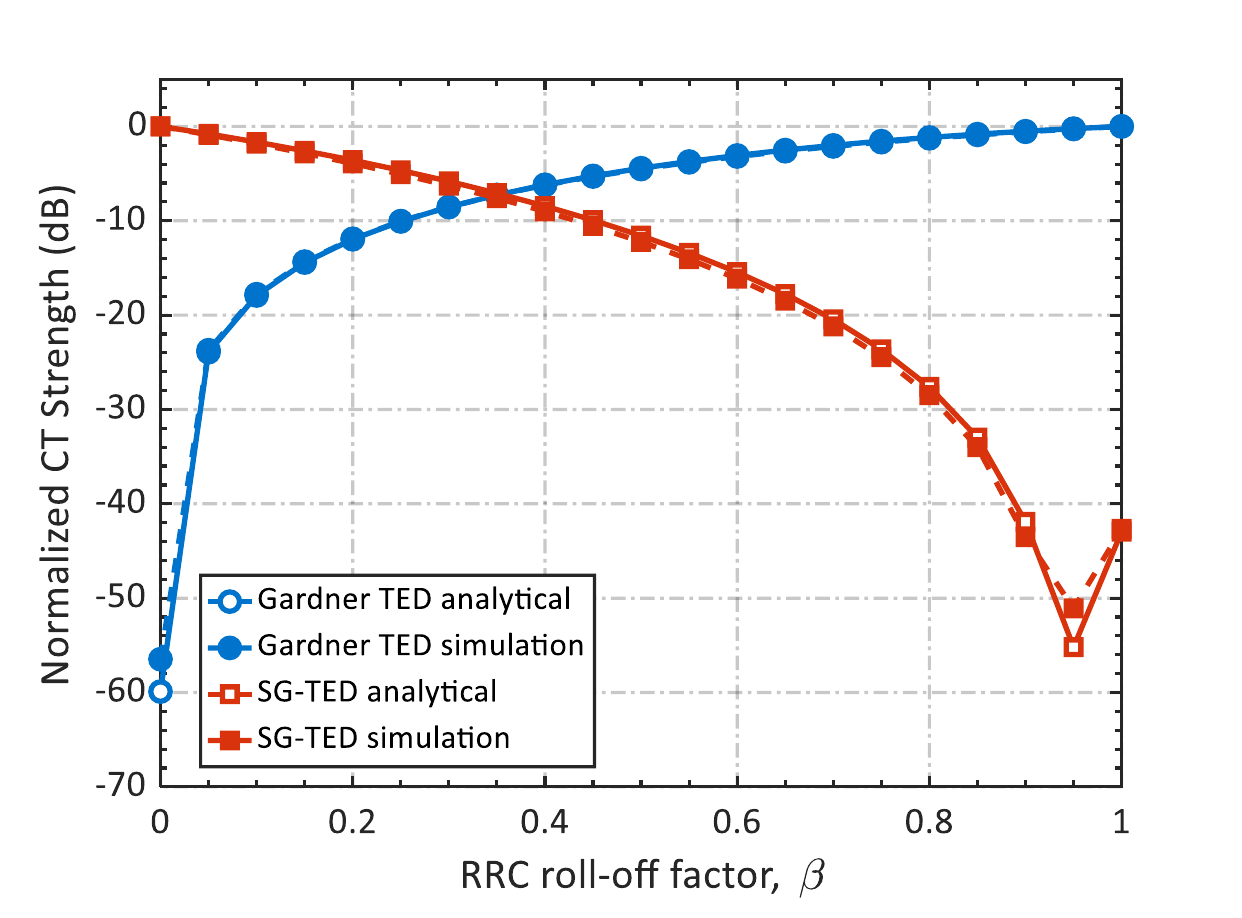}
\caption{Normalized CT strength of the Gardner TED and SG-TED as a function of the RRC roll-off factor $\beta$ for QPSK-modulated RRC-shaped signals. Solid and hollow markers denote waveform-level simulation and analytical results, respectively.}
\label{fig:word_2}
\end{figure}

We first compare the normalized CT strengths of the Gardner TED and the SG-TED for Nyquist quadrature phase-shift keying (QPSK) signals with RRC pulse shaping under different ROFs, focusing on the ROF-dependent behavior predicted by the proposed cyclic-statistical analysis. Fig.~\ref{fig:word_2} shows the normalized CT strengths of the Gardner TED and the SG-TED versus the RRC ROF, including both analytical and simulation results. The analytical and simulated curves agree well in terms of the ROF trend, validating the derived CT expressions. The small point-wise deviations are mainly attributed to finite-length and impairment effects in the Monte-Carlo simulations. For the Gardner TED, the CT strength is weak in the small-ROF region and increases monotonically as the excess bandwidth becomes larger. This trend follows from the second-order cyclic-spectral mechanism of the Gardner TED, where the symbol-rate cyclic component is determined by the overlap between the shifted pulse spectra $G(f+1/(2T_s))$ and $G(f-1/(2T_s))$. As the ROF decreases, this overlap region becomes narrower, weakening the extracted second-order CT component. In the zero-ROF limit, the CT is therefore strongly suppressed.

\begin{figure}[!t]
\centering
\includegraphics[width=1\linewidth]{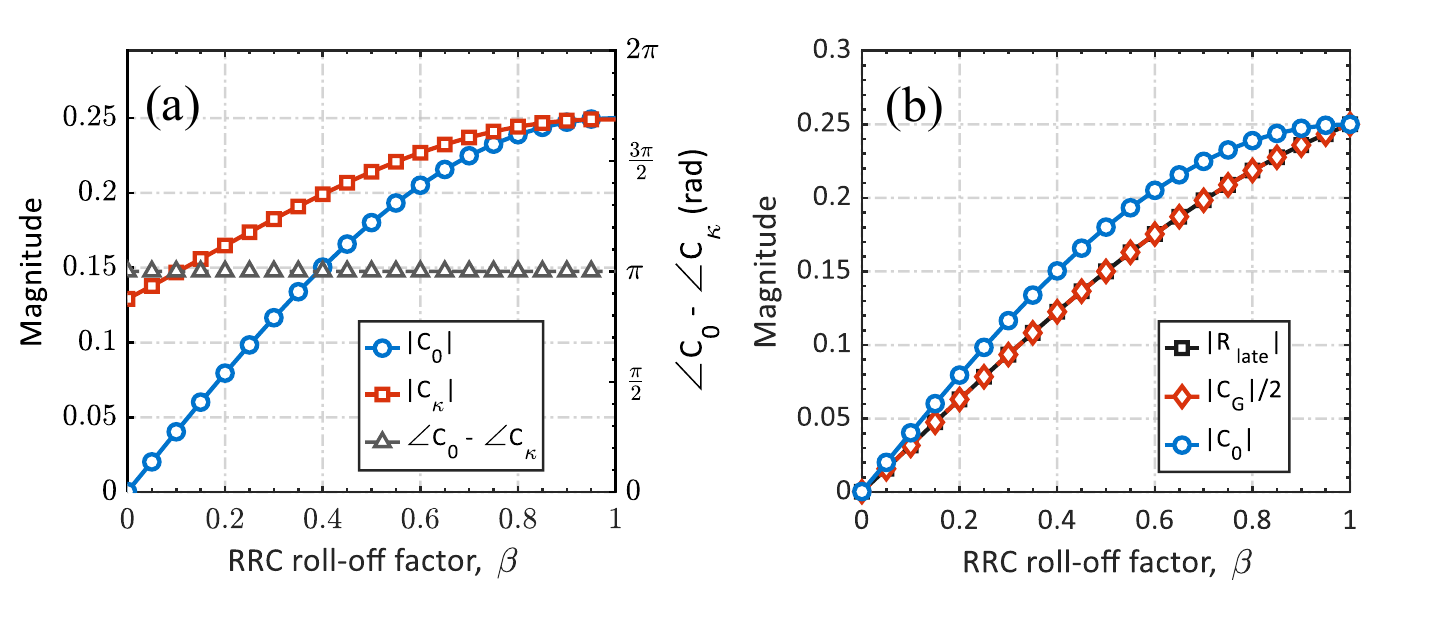}
\caption{Component-level interpretation of the SG-TED CT response under RRC pulse shaping. (a) Magnitudes of the Wick-reducible component $|C_0|$, cumulant-related component $|C_\kappa|$, and their phase difference versus the RRC roll-off factor $\beta$. (b) Comparison among $|R_{\mathrm{late}}|$, $|C_G|/2$, and $|C_0|$, illustrating the connection between the Gardner TED and the Wick-reducible component of the SG-TED.}
\label{fig:word_3}
\end{figure}

In contrast, the SG-TED exhibits an overall opposite dependence under the same conditions. Its CT strength is enhanced in the small-ROF region and decreases as the excess bandwidth increases. To clarify this behavior, Fig.~\ref{fig:word_3}(a) separately plots the magnitudes of the Wick-reducible component $|C_0|$, the cumulant-related component $|C_\kappa|$, and their phase difference.
It can be observed that the phase difference remains close to $\pi$ over the considered ROF range, indicating that the two components are nearly out of phase. For normalized QPSK, $\mathbb{E}\{|a_k|^2\}=1$ and $\mathbb{E}\{|a_k|^4\}=1$, yielding $\kappa_a=-1$. Therefore, the cumulant-related term contains a sign reversal with respect to the pulse-dependent kernel, which contributes to the nearly opposite phase relation between $C_0$ and $C_\kappa$. Consequently, the SG-TED CT strength, given by $4|C_0+C_\kappa|^2$, is mainly determined by the residual magnitude after their coherent cancellation. As the ROF increases, $|C_0|$ and $|C_\kappa|$ become closer in magnitude, strengthening their cancellation and reducing the SG-TED CT strength. This mechanism is fundamentally different from the second-order cyclic-spectral overlap governing the Gardner TED.

Fig.~\ref{fig:word_3}(b) further illustrates the connection between the conventional Gardner TED and the Wick-reducible component of the SG-TED. For the Gardner TED, the CT component is obtained from the difference between the positive- and negative-lag second-order cyclic correlations. Under the considered symmetric pulse-shaping condition, the two lag components have nearly identical magnitudes and contribute oppositely in phase, so that $|C_G|/2$ closely follows $|R_{\mathrm{late}}|$. This confirms that the factor of two in the Gardner CT amplitude originates from the early-late correlation difference. In comparison, $|C_0|$ exhibits a similar roll-off-dependent trend but does not coincide with $|R_{\mathrm{late}}|$. This is because $C_0$ is not a second-order cyclic correlation itself. Instead, it is the symbol-rate component of the squared magnitude of the same underlying field-correlation kernel. Therefore, Fig.~\ref{fig:word_3}(b) shows that the Wick-reducible term in the SG-TED is closely related to the Gardner/Godard second-order mechanism, but represents its fourth-order counterpart rather than an identical quantity.

\begin{figure}[!t]
\centering
\includegraphics[width=0.94\linewidth]{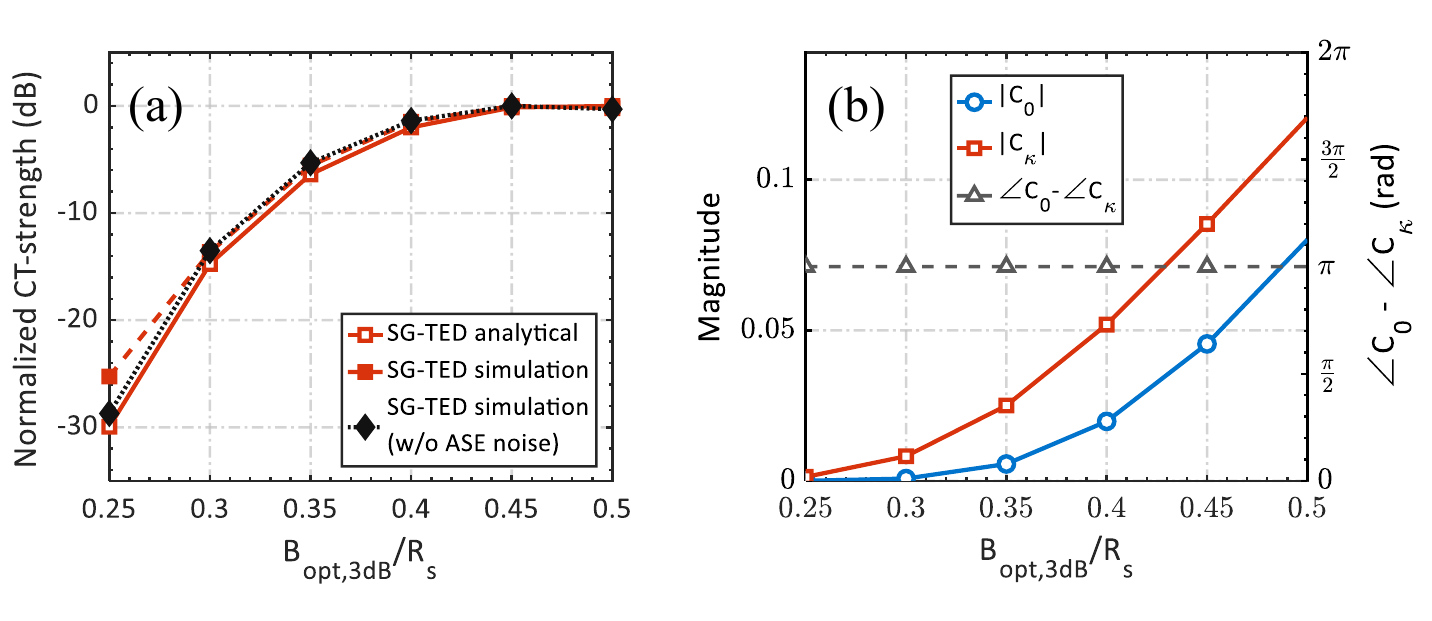}
\caption{SG-TED CT response under Gaussian pulse shaping. (a) Normalized CT strength versus the ratio between the 3-dB optical bandwidth and the symbol rate, $B_{\mathrm{opt},3\mathrm{dB}}/R_s$. Analytical results are compared with waveform-level simulations with and without ASE noise. (b) Magnitudes of the Wick-reducible component $|C_0|$, cumulant-related component $|C_\kappa|$, and their phase difference versus $B_{\mathrm{opt},3\mathrm{dB}}/R_s$.}
\label{fig:word_4}
\end{figure}

To further examine the pulse-shape dependence, Fig.~\ref{fig:word_4}(a) plots the normalized SG-TED CT strength under Gaussian pulse shaping versus the optical-bandwidth-to-symbol-rate ratio, $B_{\mathrm{opt},3\mathrm{dB}}/R_s$. A smaller $B_{\mathrm{opt},3\mathrm{dB}}/R_s$ corresponds to a broader pulse in the time domain and thus stronger temporal spreading. Both the analytical and waveform-level simulation results show that the CT strength decreases as the bandwidth becomes narrower, confirming that the derived expression characterizes the dependence of the SG-TED response on Gaussian pulse broadening. The noise-free simulation agrees more closely with the analytical curve, whereas the noisy simulation exhibits slight deviations due to ASE-induced nonlinear noise terms in the SG-TED output. Fig.~\ref{fig:word_4}(b) plots the corresponding component decomposition. Similar to the RRC case, $|C_0|$ and $|C_\kappa|$ remain nearly out of phase. As $B_{\mathrm{opt},3\mathrm{dB}}/R_s$ decreases, their residual magnitude after cancellation becomes smaller, which explains the weakened SG-TED CT in Fig.~\ref{fig:word_4}(a).

\subsection{Effects of Modulation and Symbol Distribution}

\begin{figure}[!t]
\centering
\includegraphics[width=1\linewidth]{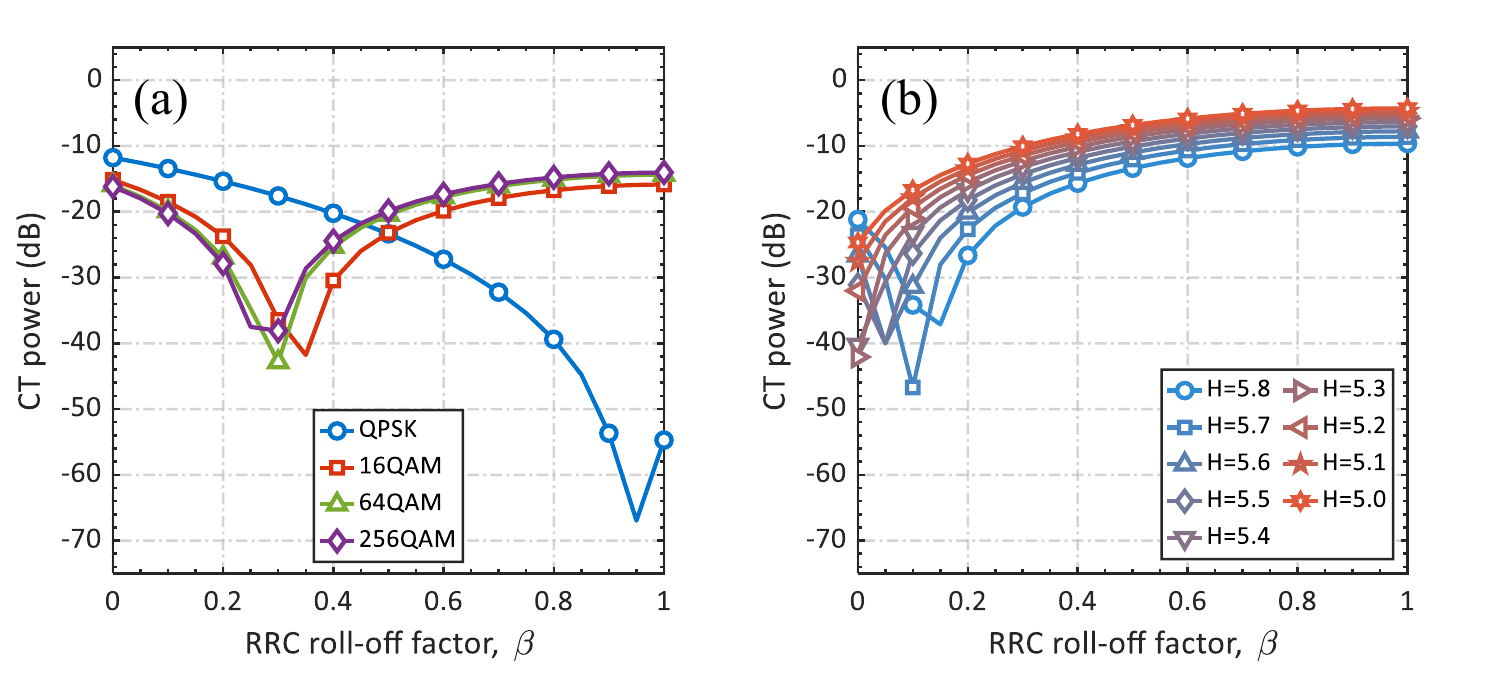}
\caption{SG-TED CT power under different modulation formats and symbol distributions with RRC pulse shaping. (a) Uniform QAM formats, including QPSK, 16QAM, 64QAM, and 256QAM. (b) PS-64QAM signals with shaping entropy $H$ varying from 5.8 to 5.0 bits/symbol.}
\label{fig:word_5}
\end{figure}

We next evaluate the effect of modulation format and symbol distribution on the SG-TED CT response under RRC pulse shaping. Fig.~\ref{fig:word_5}(a) shows the CT strength versus the ROF for QPSK, 16QAM, 64QAM, and 256QAM. Different modulation formats exhibit distinct ROF-dependent responses, indicating that the SG-TED is sensitive to modulation-dependent fourth-order statistics. Fig.~\ref{fig:word_5}(b) further shows the results for probabilistically shaped 64QAM (PS-64QAM), where the entropy $H$ is varied from 5.8 to 5 bits/symbol. As the shaping entropy decreases, the CT strength curve changes noticeably, since probabilistic shaping modifies the symbol distribution and hence the fourth-order cumulant $\kappa_a$. These results confirm that the SG-TED CT strength is determined not only by the pulse shape, but also by the higher-order statistical properties of the transmitted symbols.

\begin{figure}[!t]
\centering
\includegraphics[width=1\linewidth]{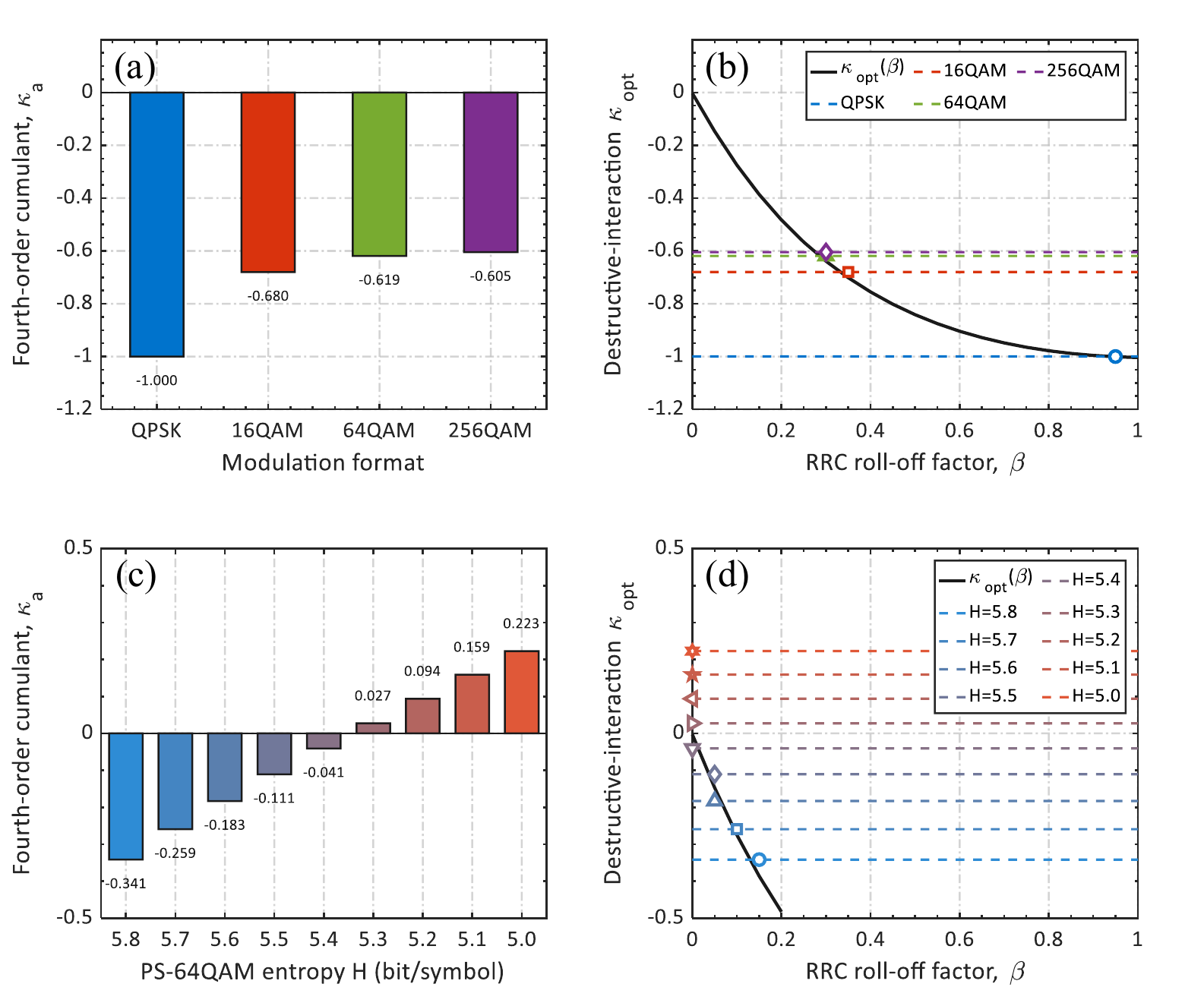}
\caption{Modulation-dependent fourth-order cumulant and pulse-dependent CT cancellation condition of the SG-TED. (a) Calculated $\kappa_a$ values for uniform QAM formats. (b) Intersection between the QAM-dependent $\kappa_a$ values and the pulse-dependent optimal cancellation condition $\kappa_{\mathrm{opt}}(\beta)$. (c) Calculated $\kappa_a$ values for PS-64QAM with entropy $H$ from 5.8 to 5.0 bits/symbol. (d) Comparison between PS-64QAM-dependent $\kappa_a$ and $\kappa_{\mathrm{opt}}(\beta)$.}
\label{fig:word_6}
\end{figure}

The modulation-dependent minima and turning points observed in Fig.~\ref{fig:word_5} can be explained by comparing $\kappa_a$ with the pulse-dependent cancellation condition. Fig.~\ref{fig:word_6}(a) shows the calculated $\kappa_a$ values for different normalized QAM formats. The cumulant is negative for all considered uniform QAM signals, and its magnitude decreases as the modulation order increases, i.e., $\kappa_a$ gradually moves toward zero. Fig.~\ref{fig:word_6}(b) plots the corresponding optimal cancellation condition $\kappa_{\mathrm{opt}}(\beta)$, which is obtained for each ROF by minimizing the CT component,

\begin{equation}
\label{eq:word_61}
\kappa_{\mathrm{opt}}(\beta)
=\underset{\kappa\in\mathbb{R}}{\arg\min}\,
\left|C_0(\beta)+\kappa D_\beta^{1/T_s}\!\left(T_s/2\right)\right|^2.
\end{equation}

The black curve represents the pulse-dependent optimal cumulant, while the dashed horizontal lines denote the cumulants of different modulation formats. Therefore, the intersection between a dashed line and the black curve indicates the ROF at which $C_0$ and $C_\kappa$ exhibit the strongest cancellation, corresponding to the CT-strength minimum in Fig.~\ref{fig:word_5}(a). The markers indicate the closest ROF points on the finite plotting grid. This comparison shows that the CT minima originate from the matching between the symbol-dependent cumulant and the pulse-dependent cancellation condition.

Fig.~\ref{fig:word_6}(c) shows the variation of $\kappa_a$ for PS-64QAM as the shaping entropy decreases from 5.8 to 5.0 bits/symbol. With stronger probabilistic shaping, $\kappa_a$ increases from a negative value and crosses zero between $H=5.4$ and $H=5.3$. This sign change modifies the contribution of $C_\kappa$ to the coherent sum $C_0+C_\kappa$. When $\kappa_a<0$, the cumulant-related component tends to cancel the Wick-reducible component under the considered pulse conditions. When $\kappa_a>0$, this cancellation is weakened or becomes constructive, leading to an enhanced CT response. Fig.~\ref{fig:word_6}(d) further confirms this interpretation: only weakly shaped PS-64QAM with negative $\kappa_a$ can approach the cancellation condition, whereas stronger shaping moves $\kappa_a$ away from the negative-cancellation region.

\begin{figure}[!t]
\centering
\includegraphics[width=1\linewidth]{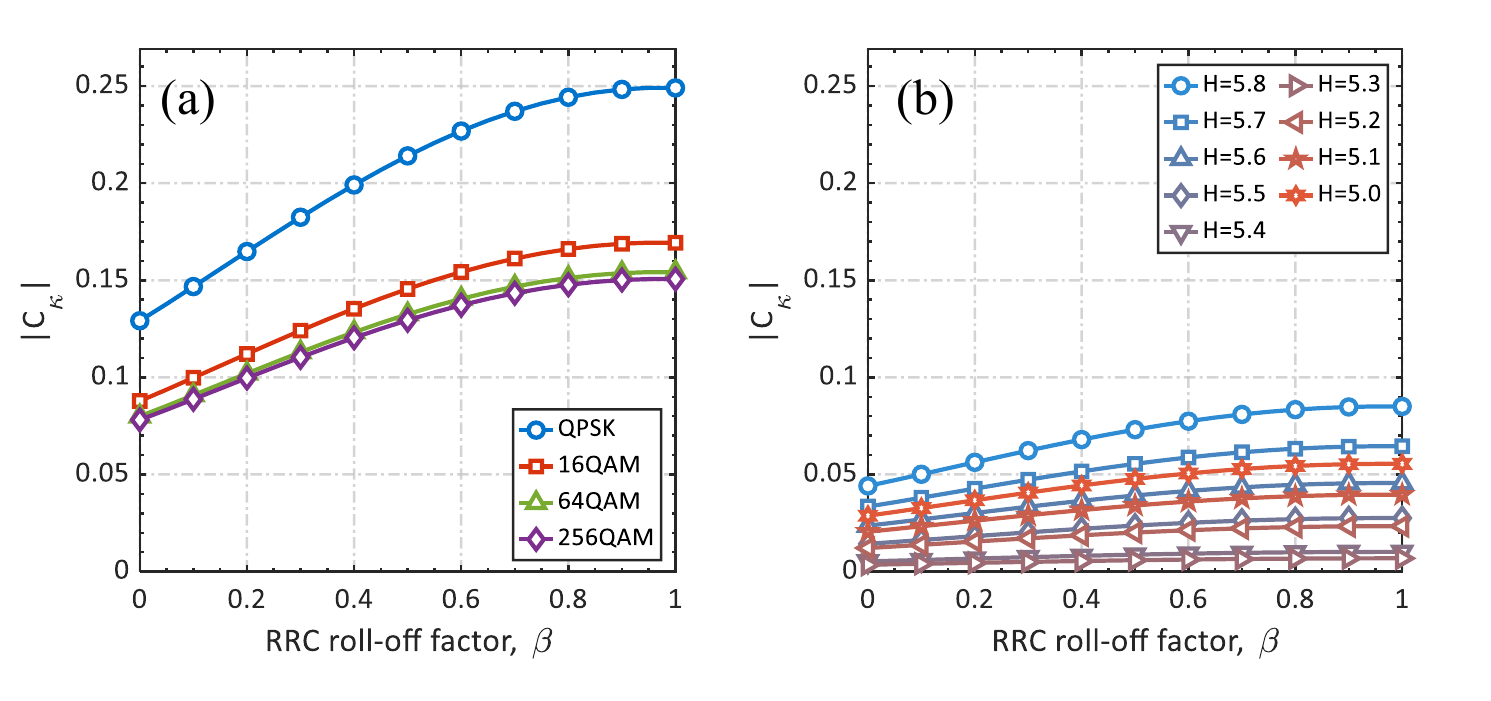}
\caption{Modulation- and distribution-dependent cumulant-related component of the SG-TED. (a) $|C_\kappa|$ for uniform QAM formats versus the RRC ROF $\beta$. (b) $|C_\kappa|$ for PS-64QAM signals with different shaping entropies versus $\beta$.}
\label{fig:word_7}
\end{figure}

Fig.~\ref{fig:word_7}(a) and Fig.~\ref{fig:word_7}(b) further plot the magnitudes of the cumulant-related component $|C_\kappa|$ for uniform QAM formats and PS-64QAM distributions, respectively. Since the Wick-reducible component $C_0$ is mainly determined by the pulse shape, the modulation- and distribution-dependent behavior of the SG-TED is primarily introduced through $C_\kappa$. For uniform QAM signals, QPSK produces the largest $|C_\kappa|$ owing to its largest negative fourth-order cumulant, whereas higher-order QAM formats exhibit relatively smaller and closer $|C_\kappa|$ levels. In contrast, PS-64QAM generally yields a much weaker cumulant-related contribution. This is because probabilistic shaping reduces the magnitude of $\kappa_a$ and can even change its sign as the shaping entropy decreases.

As shown in Fig.~\ref{fig:word_7}(b), when the entropy decreases, $|C_\kappa|$ is first weakened as $\kappa_a$ approaches zero, and then increases again after $\kappa_a$ becomes positive. However, compared with uniform low-order QAM, especially QPSK, the cumulant-related contribution of PS-64QAM remains significantly weaker over a wide entropy range. Since the SG-TED relies on the coherent superposition of the Wick-reducible component and the cumulant-related non-Gaussian component, a weakened or sign-changed $C_\kappa$ can substantially reduce the available CT. This provides a fourth-order statistical explanation for the degraded or even failed timing recovery of power-based Gardner schemes for PS-QAM signals.

\subsection{Effects of Polarization Rotation and PMD}

\begin{figure}[!t]
\centering
\includegraphics[width=1\linewidth]{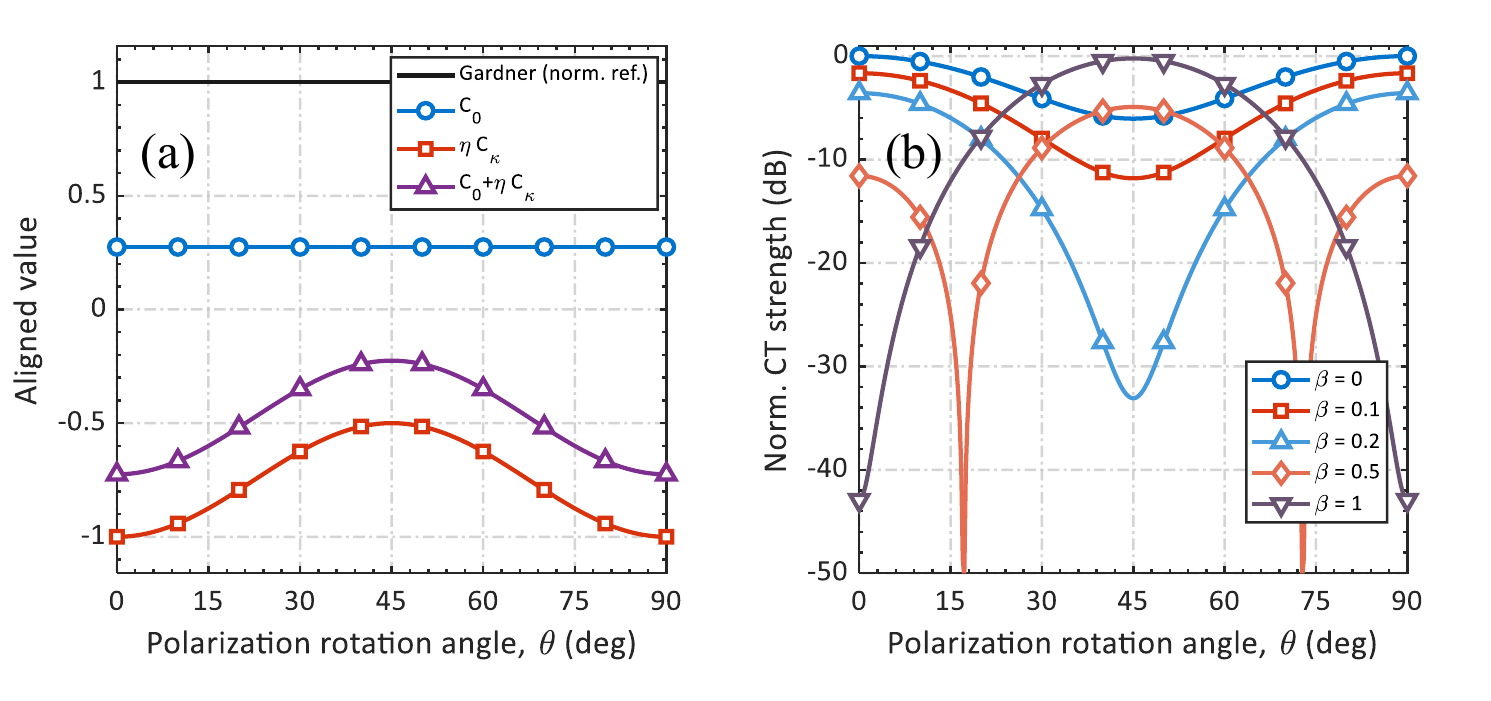}
\caption{Effect of static polarization rotation on the Gardner TED and SG-TED under RRC pulse shaping. (a) Polarization-rotation dependence of Gardner, the Wick-reducible component $C_0$, the scaled cumulant-related component $\eta C_\kappa$, and their coherent sum $C_0+\eta C_\kappa$ for $\beta=0.1$. (b) Normalized SG-TED CT strength versus the polarization rotation angle for different RRC ROFs.}
\label{fig:word_8}
\end{figure}

We then investigate the CT responses under static polarization rotation and first-order PMD. Fig.~\ref{fig:word_8}(a) shows the polarization-rotation dependence of CT components for an RRC-shaped QPSK signal with an ROF of $\beta=0.1$. According to Eqs.~\eqref{eq:word_46} and \eqref{eq:word_47}, static polarization rotation does not change the second-order cyclic-statistical response under identical and independent dual-polarization tributaries. Therefore, the Gardner TED remains insensitive to the polarization rotation angle, as shown by the normalized reference curve. Similarly, the Wick-reducible component $C_0$ of the SG-TED is also independent of the rotation angle, since it is determined by the second-order field-correlation structure. In contrast, the cumulant-related component is scaled by $\eta(\theta)=1-0.5\sin^2(2\theta)$, which reaches its minimum at $\theta=45^\circ$ and its periodic equivalents.

\begin{figure}[!t]
\centering
\includegraphics[width=0.9\linewidth]{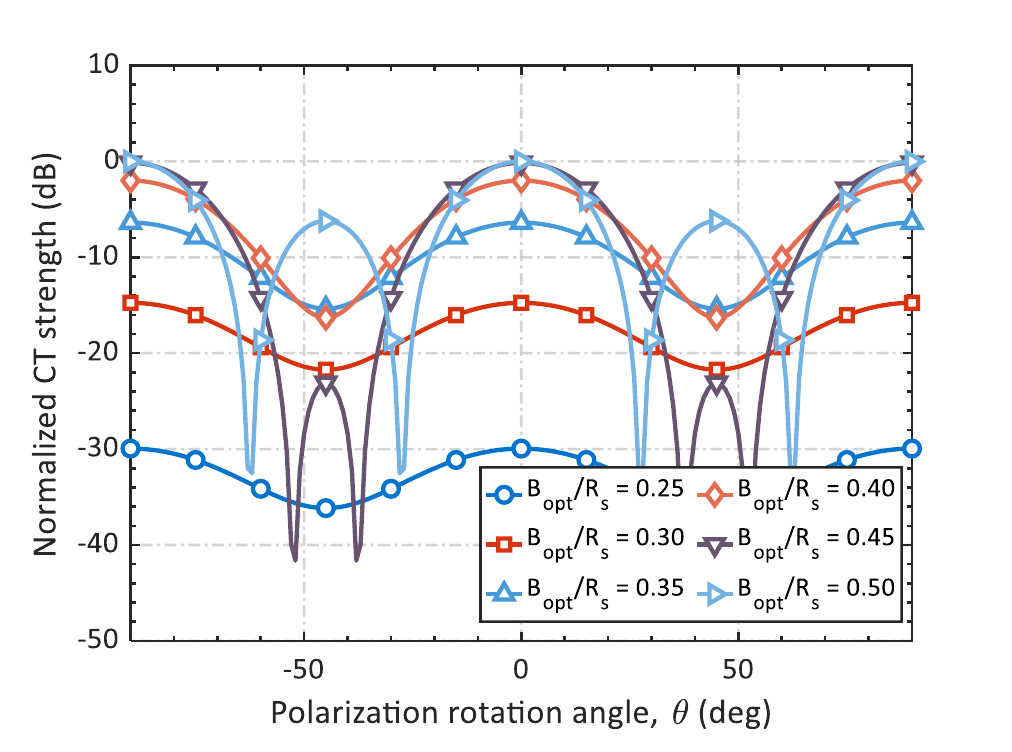}
\caption{Static-polarization-rotation response of the SG-TED under Gaussian pulse shaping. Normalized CT strength versus the polarization rotation angle for different ratios between the 3-dB optical bandwidth and the symbol rate.}
\label{fig:word_9}
\end{figure}

Fig.~\ref{fig:word_8}(b) further plots the normalized SG-TED CT strength versus the polarization rotation angle for different RRC ROFs. Although $\eta(\theta)$ itself is minimized at $\theta=45^\circ$, the minimum CT strength does not necessarily occur at this angle. This is because the observable CT strength is proportional to $|C_0+\eta(\theta)C_\kappa|^2$. When $C_0$ and $C_\kappa$ are nearly out of phase, the strongest fading occurs when the scaled cumulant-related term $\eta(\theta)C_\kappa$ best matches $C_0$ in magnitude, rather than when $\eta(\theta)$ is simply minimized. As the ROF increases, the relative magnitudes of $C_0$ and $C_\kappa$ change, causing the strongest cancellation angle to shift away from $\theta=45^\circ$. In particular, for $\beta=1$, $|C_0|$ and $|C_\kappa|$ are close in magnitude. In this case, the strongest cancellation occurs near $\eta(\theta)\approx 1$, corresponding to angles close to $0^\circ$ or $90^\circ$. At $\theta=45^\circ$, however, $\eta(\theta)$ is reduced, so the cumulant-related term is weakened and the cancellation becomes less effective. As a result, $\theta=45^\circ$ becomes a relatively strong CT-response position rather than a fading point. 

Fig.~\ref{fig:word_9} shows the polarization-rotation-dependent SG-TED CT strength under Gaussian pulse shaping with different $B_{\mathrm{opt},3\mathrm{dB}}/R_s$. Changing this ratio modifies the Gaussian pulse width and hence the relative balance between $C_0$ and $C_\kappa$, resulting in different polarization-dependent fading angles and depths. This behavior is consistent with the RRC case and confirms that the SG-TED response is determined by the coherent balance between $C_0$ and $\eta(\theta)C_\kappa$, rather than by $\eta(\theta)$ alone. Therefore, practical SG-TED-based timing recovery requires polarization correction to mitigate polarization-dependent CT fading \cite{ref22,ref23}.

\begin{figure*}[!t]
\centering
\includegraphics[width=0.85\textwidth]{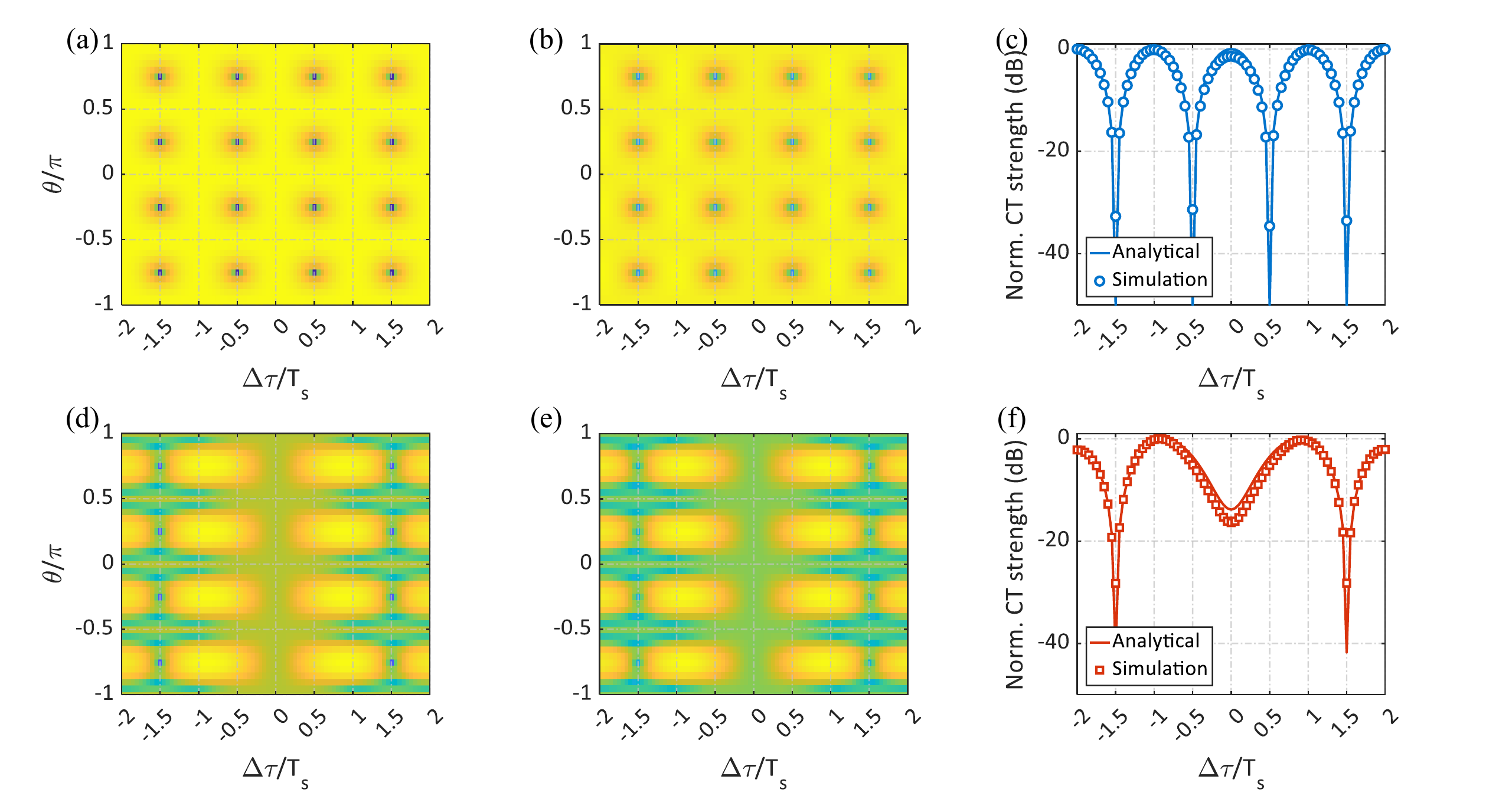}
\caption{Analytical and waveform-level simulation validation of PMD-induced CT fading for RRC pulse shaping with $\beta=0.5$. (a) Analytical and (b) simulated CT-strength maps of the Gardner TED under first-order PMD. (c) One-dimensional cut comparing the analytical and simulated Gardner TED results. (d) Analytical and (e) simulated CT-strength maps of the SG-TED under first-order PMD. (f) One-dimensional cut comparing the analytical and simulated SG-TED results.}
\label{fig:word_10}
\end{figure*}

To evaluate the PMD-induced CT-strength variation, we further characterize the SG-TED CT response under first-order PMD and validate the analytical PMD expressions using representative waveform-level simulations. Fig.~\ref{fig:word_10} compares the analytical and simulated results for an RRC-shaped signal with $\beta=0.5$ under first-order PMD. Fig.~\ref{fig:word_10}(a) and Fig.~\ref{fig:word_10}(b) show the analytical and simulated CT-strength maps of the Gardner TED, respectively, while Fig.~\ref{fig:word_10}(c) gives a representative one-dimensional cut for direct comparison. The simulation reproduces the main fading locations predicted by the analytical model. Deep CT fading occurs when the polarization rotation angle approaches $45^\circ$ and its periodic equivalents, together with a normalized DGD close to odd multiples of half the symbol period. This behavior is consistent with the second-order PMD fading mechanism.

Fig.~\ref{fig:word_10}(d) and Fig.~\ref{fig:word_10}(e) further compare the analytical and simulated CT-strength maps of the SG-TED, and Fig.~\ref{fig:word_10}(f) shows the corresponding one-dimensional cut. Compared with the Gardner TED, the SG-TED exhibits a more complicated PMD-dependent fading pattern. This is because its CT strength is determined by the coherent superposition of the Wick-reducible component and the cumulant-related component, rather than by a single second-order PMD fading factor. Therefore, the PMD response is reshaped by the relative amplitudes and phases of the fourth-order components. Overall, the analytical and simulated results agree well in terms of the fading locations and overall CT-strength variation.

\begin{figure}[!t]
\centering
\includegraphics[width=0.95\linewidth]{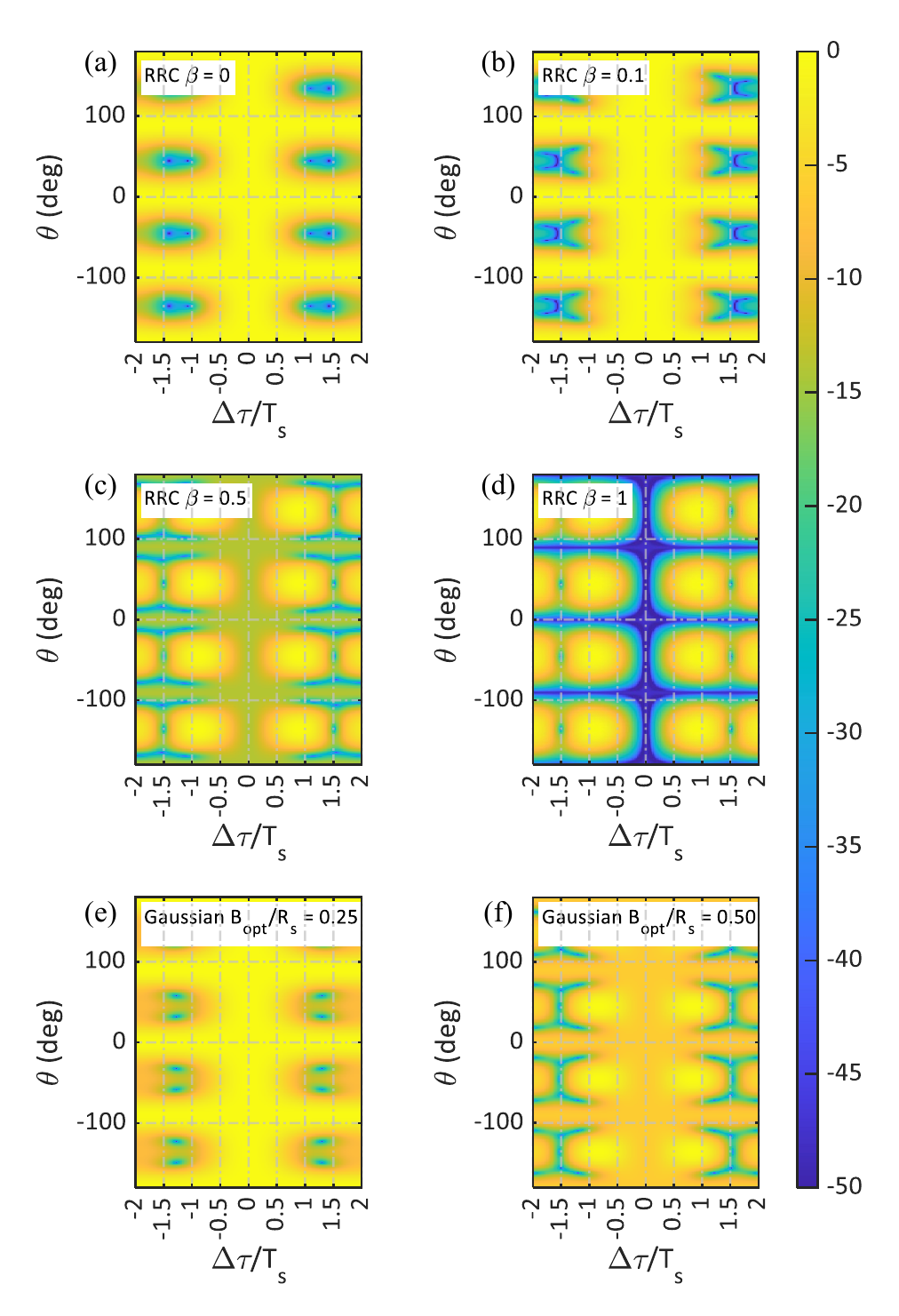}
\caption{Analytical SG-TED CT-strength maps under first-order PMD for different pulse-shaping conditions. (a)--(d) RRC pulse shaping with roll-off factors of $\beta=0$, 0.1, 0.5, and 1, respectively. (e), (f) Gaussian pulse shaping with $B_{\mathrm{opt},3\mathrm{dB}}/R_s=0.25$ and 0.5, respectively.}
\label{fig:word_11}
\end{figure}

Fig.~\ref{fig:word_11} then evaluates the PMD-dependent CT strength of the SG-TED under different pulse-shaping conditions. Six representative cases are considered, including RRC pulses with ROFs of $\beta=0$, 0.1, 0.5, and 1, and Gaussian pulses with $B_{\mathrm{opt},3\mathrm{dB}}/R_s=0.25$ and 0.5. The results show that the PMD-induced fading behavior of the SG-TED strongly depends on the pulse shape and pulse bandwidth. For RRC pulse shaping, changing the ROF significantly modifies the fading locations and fading depths. In particular, as $\beta$ increases, the CT-strength distribution evolves from isolated fading regions to more extended fading bands. This indicates that the PMD response of the SG-TED is not governed by a fixed Gardner-type nulling condition. Instead, it is determined by the coherent superposition of the PMD-modified Wick-reducible and cumulant-related components. For Gaussian pulse shaping, changing the 3-dB optical-bandwidth-to-symbol-rate ratio also leads to different CT-fading patterns. A narrower optical bandwidth broadens the pulse in time and changes the relative balance between the fourth-order components, thereby modifying the PMD sensitivity of the SG-TED. Overall, Fig.~\ref{fig:word_11} confirms that the SG-TED exhibits pulse-dependent and fourth-order-statistics-dependent PMD fading, which is fundamentally different from the simpler second-order PMD behavior of the Gardner TED.

\subsection{Other Power-Based Modified Gardner TEDs}

The above fourth-order cyclic-statistical analysis can also provide a useful reference for other power-based modified Gardner TEDs, such as the IP-Gardner \cite{ref25} and NIP-Gardner \cite{ref8} schemes. Although these TEDs apply different nonlinear transformations to the sampling points involved in the original Gardner TED, their timing information is still generated from nonlinear intensity-dependent functions of the received waveform. Therefore, their CT generation can be analyzed in a similar way: first expressing the TED output as a nonlinear functional of the sampled field or power waveform, then extracting its symbol-rate cyclic component, and finally applying moment and cumulant decompositions to separate the Wick-reducible contribution from the cumulant-related non-Gaussian contribution.

Compared with the SG-TED, the interpolation and normalization operations in IP-Gardner and NIP-Gardner will modify the pulse-dependent kernels and scaling factors, but they do not change the essential fact that the timing tone is governed by higher-order cyclic statistics. Therefore, the analytical framework developed in this work can be extended to evaluate how pulse shaping, modulation format, symbol distribution, polarization rotation, and PMD affect these modified Gardner TEDs. A complete derivation for each variant is beyond the scope of this paper, but the present analysis establishes a general statistical methodology for understanding power-based Gardner-type timing recovery algorithms.

\section{Conclusions}

In this paper, we developed a fourth-order cyclostationary analytical framework for power-based nonlinear Gardner TEDs, with the SG-TED taken as a representative case. The analysis revealed that the SG-TED CT originates from the symbol-rate cyclic component of the power-process ACF and is therefore governed by fourth-order cyclic statistics of the received complex field. By decomposing the CT component into Wick-reducible and cumulant-related non-Gaussian terms, we clarified the connection between the SG-TED and the conventional Gardner/Godard mechanism, as well as the origin of its modulation- and distribution-dependent behavior. The effects of pulse shaping, probabilistic shaping, static polarization rotation, and first-order PMD were systematically analyzed, showing that the SG-TED exhibits CT enhancement and fading behaviors fundamentally different from second-order TEDs. Numerical evaluations and waveform-level simulations verified the analytical predictions. The proposed framework provides a unified statistical basis for understanding SG-TED and related power-based Gardner-type TEDs, including IP-Gardner and NIP-Gardner.

\appendices
\makeatletter
\@addtoreset{equation}{section}
\makeatother
\renewcommand{\theequation}{\Alph{section}\arabic{equation}}
\section{Derivation of the SG-TED Cyclic Component}
\label{app:sg_cyclic}

Define the symmetric power-correlation kernel as

\begin{equation}
\label{eq:word_62}
K_p(t,u)=\mathbb{E}\!\left[
p\!\left(t+\frac{u}{2}\right)p\!\left(t-\frac{u}{2}\right)
\right].
\end{equation}
Using $u_0=T_s/2$, the two terms in the SG-TED output satisfy

\begin{equation}
\label{eq:word_63}
\begin{aligned}
\mathbb{E}\!\left[p(t)p^*(t-u_0)\right]
&=K_p\!\left(t-\frac{u_0}{2},u_0\right),\\
\mathbb{E}\!\left[p(t)p^*(t+u_0)\right]
&=K_p\!\left(t+\frac{u_0}{2},u_0\right).
\end{aligned}
\end{equation}
Therefore,

\begin{equation}
\label{eq:word_64}
\mathbb{E}\!\left[\varepsilon_{\mathrm{SG}}(t)\right]
=K_p\!\left(t-\frac{u_0}{2},u_0\right)
-K_p\!\left(t+\frac{u_0}{2},u_0\right).
\end{equation}
The cyclic coefficient of $K_p(t,u)$ at cyclic frequency $\alpha$ is $R_p^\alpha(u)$. By the time-shift property of Fourier coefficients,

\begin{equation}
\label{eq:word_65}
\begin{aligned}
\left[K_p\!\left(t-\frac{u_0}{2},u_0\right)\right]_\alpha
&=e^{-j\pi\alpha u_0}R_p^\alpha(u_0),\\
\left[K_p\!\left(t+\frac{u_0}{2},u_0\right)\right]_\alpha
&=e^{j\pi\alpha u_0}R_p^\alpha(u_0).
\end{aligned}
\end{equation}
Thus,

\begin{equation}
\label{eq:word_66}
\begin{aligned}
C_{\mathrm{SG}}^\alpha(u_0)
&=\left(e^{-j\pi\alpha u_0}-e^{j\pi\alpha u_0}\right)R_p^\alpha(u_0)\\
&=-2j\sin(\pi\alpha u_0)R_p^\alpha(u_0).
\end{aligned}
\end{equation}

\section{Fourth-Order Moment Expansion of the Power Autocorrelation}
\label{app:fourth_moment}

Using Eq.~\eqref{eq:word_22}, we have

\begin{equation}
\label{eq:word_67}
\mathbb{E}\!\left[|x(t_+)|^2|x(t_-)|^2\right]
=\sum_{k,l,m,n}\mathbb{E}\!\left[a_ka_l^*a_ma_n^*\right]
g_{k,+}g_{l,+}^*g_{m,-}g_{n,-}^*.
\end{equation}
For independent zero-mean proper complex symbols,

\begin{equation}
\label{eq:word_68}
\mathbb{E}\!\left[a_ka_l^*\right]=\sigma_a^2\delta_{kl},\qquad
\mathbb{E}\!\left[a_ka_m\right]=0.
\end{equation}
The fourth-order moment can be

\begin{equation}
\label{eq:word_69}
\mathbb{E}\!\left[a_ka_l^*a_ma_n^*\right]
=\sigma_a^4\delta_{kl}\delta_{mn}
+\sigma_a^4\delta_{kn}\delta_{ml}
+\kappa_a\delta_{klmn},
\end{equation}
where

\begin{equation}
\label{eq:word_70}
\kappa_a=\mathbb{E}\!\left[|a_k|^4\right]-2\sigma_a^4.
\end{equation}
Substituting Eqs.~\eqref{eq:word_68} and \eqref{eq:word_69} into Eq.~\eqref{eq:word_67} gives

\begin{equation}
\label{eq:word_71}
\begin{aligned}
\mathbb{E}\!\left[|x(t_+)|^2|x(t_-)|^2\right]
=\sigma_a^4\sum_k |g_{k,+}|^2\sum_m |g_{m,-}|^2\\
+\sigma_a^4\left|\sum_k g_{k,+}g_{k,-}^*\right|^2+\kappa_a\sum_k |g_{k,+}|^2|g_{k,-}|^2.
\end{aligned}
\end{equation}
Using the definitions of $m(t)$, $w_x(t,T_s/2)$, and $d(t,T_s/2)$, Eq.~\eqref{eq:word_71} reduces to

\begin{equation}
\label{eq:word_72}
\begin{aligned}
\mathbb{E}\!\left[|x(t_+)|^2|x(t_-)|^2\right]
&=m\!\left(t+\frac{u}{2}\right)m\!\left(t-\frac{u}{2}\right)\\
&\quad+\left|w_x(t,u)\right|^2+\kappa_a d(t,u).
\end{aligned}
\end{equation}

\section{Proof of the Vanishing Average-Power Term}
\label{app:vanishing_average}

For $u_0=T_s/2$, the average-power product is

\begin{equation}
\label{eq:word_73}
F_m(t)=m\!\left(t+\frac{T_s}{4}\right)
m\!\left(t-\frac{T_s}{4}\right).
\end{equation}
Since $m(t)$ is $T_s$-periodic, $m(t+T_s)=m(t)$. Therefore,

\begin{equation}
\label{eq:word_77}
\begin{aligned}
F_m\!\left(t+\frac{T_s}{2}\right)
&=m\!\left(t+\frac{3T_s}{4}\right)
m\!\left(t+\frac{T_s}{4}\right)\\
&=m\!\left(t-\frac{T_s}{4}\right)
m\!\left(t+\frac{T_s}{4}\right)=F_m(t).
\end{aligned}
\end{equation}
Thus, $F_m(t)$ has a period of $T_s/2$, and its Fourier series contains only even harmonics of $1/T_s$. Consequently, its symbol-rate Fourier coefficient is zero,

\begin{equation}
\label{eq:word_78}
M^{1/T_s}(T_s/2)=0.
\end{equation}

\section{PMD-Distorted Cyclic Components}
\label{app:pmd}
\subsection{Second-Order PMD Response}
\label{app:pmd_second_order}

Let the transmitted and received dual-polarization frequency domain vectors be denoted by

\begin{equation}
\label{eq:word_79}
\mathbf{T}(f)=
\begin{bmatrix}T_x(f)\\ T_y(f)\end{bmatrix},\qquad
\mathbf{R}(f)=
\begin{bmatrix}R_x(f)\\ R_y(f)\end{bmatrix}.
\end{equation}
For a first-order PMD channel, we have

\begin{equation}
\label{eq:word_80}
\mathbf{R}(f)=\mathbf{H}(f)\mathbf{T}(f),
\end{equation}
where

\begin{equation}
\label{eq:word_81}
\mathbf{H}(f)=\mathbf{U}(\theta,\phi)
\begin{bmatrix}
e^{-j\pi f\Delta\tau} & 0\\
0 & e^{j\pi f\Delta\tau}
\end{bmatrix}
\mathbf{U}^{H}(\theta,\phi).
\end{equation}
and

\begin{equation}
\label{eq:word_82}
\mathbf{U}(\theta,\phi)=
\begin{bmatrix}
c & -e^{j\phi}s\\
e^{-j\phi}s & c
\end{bmatrix},\qquad
c=\cos\theta,\quad s=\sin\theta.
\end{equation}
The input and output SCF matrices are defined as

\begin{equation}
\label{eq:word_83}
\begin{aligned}
\mathbf{S}_{\mathrm{in}}^\alpha(f)
&=\mathbb{E}\!\left[
\mathbf{T}\!\left(f+\frac{\alpha}{2}\right)
\mathbf{T}^{H}\!\left(f-\frac{\alpha}{2}\right)
\right],\\
\mathbf{S}_{\mathrm{out}}^\alpha(f)
&=\mathbb{E}\!\left[
\mathbf{R}\!\left(f+\frac{\alpha}{2}\right)
\mathbf{R}^{H}\!\left(f-\frac{\alpha}{2}\right)
\right].
\end{aligned}
\end{equation}
Substituting Eq.~\eqref{eq:word_80} into Eq.~\eqref{eq:word_83} gives

\begin{equation}
\label{eq:word_85}
\begin{aligned}
\mathbf{S}_{\mathrm{out}}^\alpha(f)
&=\mathbb{E}\!\left[
\mathbf{H}\!\left(f+\frac{\alpha}{2}\right)
\mathbf{T}\!\left(f+\frac{\alpha}{2}\right)
\mathbf{T}^{H}\!\left(f-\frac{\alpha}{2}\right)
\mathbf{H}^{H}\!\left(f-\frac{\alpha}{2}\right)
\right]\\
&=\mathbf{H}\!\left(f+\frac{\alpha}{2}\right)
\mathbf{S}_{\mathrm{in}}^\alpha(f)
\mathbf{H}^{H}\!\left(f-\frac{\alpha}{2}\right).
\end{aligned}
\end{equation}
where the PMD matrix is deterministic and can therefore be taken outside the expectation. Since the two input polarization tributaries are assumed to be independent and identically distributed,

\begin{equation}
\label{eq:word_86}
\mathbf{S}_{\mathrm{in}}^\alpha(f)=S_x^\alpha(f)\mathbf{I}.
\end{equation}
Thus,

\begin{equation}
\label{eq:word_87}
\mathbf{S}_{\mathrm{out}}^\alpha(f)
=S_x^\alpha(f)
\mathbf{H}\!\left(f+\frac{\alpha}{2}\right)
\mathbf{H}^{H}\!\left(f-\frac{\alpha}{2}\right).
\end{equation}

Using Eq.~\eqref{eq:word_81}, the matrix product in Eq.~\eqref{eq:word_87} becomes

\begin{equation}
\label{eq:word_88}
\mathbf{H}\!\left(f+\frac{\alpha}{2}\right)
\mathbf{H}^{H}\!\left(f-\frac{\alpha}{2}\right)
=\mathbf{U}(\theta,\phi)\boldsymbol{\Lambda}_{\alpha}\mathbf{U}^{H}(\theta,\phi).
\end{equation}
where

\begin{equation}
\label{eq:word_89}
\boldsymbol{\Lambda}_{\alpha}=
\begin{bmatrix}
e^{-j\pi\alpha\Delta\tau} & 0\\
0 & e^{j\pi\alpha\Delta\tau}
\end{bmatrix}.
\end{equation}
If only the first output polarization is observed, its cyclic spectrum is obtained by projecting $\mathbf{S}_{\mathrm{out}}^\alpha(f)$ onto the first polarization basis vector $\mathbf{e}_1=[1\ 0]^T$. Therefore,

\begin{equation}
\label{eq:word_90}
S_{x,\mathrm{PMD}}^\alpha(f)
=\mathbf{e}_1^{T}\mathbf{S}_{\mathrm{out}}^\alpha(f)\mathbf{e}_1.
\end{equation}
Substituting Eqs.~\eqref{eq:word_86} and \eqref{eq:word_87} into Eq.~\eqref{eq:word_90}, we have

\begin{equation}
\label{eq:word_91}
S_{x,\mathrm{PMD}}^\alpha(f)
=S_x^\alpha(f)\mathbf{e}_1^{T}
\mathbf{U}(\theta,\phi)\boldsymbol{\Lambda}_{\alpha}
\mathbf{U}^{H}(\theta,\phi)\mathbf{e}_1.
\end{equation}
The scalar projection in Eq.~\eqref{eq:word_91} can be calculated explicitly as

\begin{equation}
\label{eq:word_92}
\begin{aligned}
&\mathbf{e}_1^{T}\mathbf{U}(\theta,\phi)
\boldsymbol{\Lambda}_{\alpha}\mathbf{U}^{H}(\theta,\phi)\mathbf{e}_1\\
&=\begin{bmatrix}c&-e^{j\phi}s\end{bmatrix}
\begin{bmatrix}
e^{-j\pi\alpha\Delta\tau} & 0\\
0 & e^{j\pi\alpha\Delta\tau}
\end{bmatrix}
\begin{bmatrix}c\\ -e^{-j\phi}s\end{bmatrix}\\
&=c^2e^{-j\pi\alpha\Delta\tau}
+s^2e^{j\pi\alpha\Delta\tau}.
\end{aligned}
\end{equation}
Consequently, the cyclic spectrum observed at the first output branch is

\begin{equation}
\label{eq:word_93}
S_{x,\mathrm{PMD}}^\alpha(f)
=\eta_2(\alpha,\theta,\Delta\tau)S_x^\alpha(f),
\end{equation}
where

\begin{equation}
\label{eq:word_94}
\eta_2(\alpha,\theta,\Delta\tau)
=c^2e^{-j\pi\alpha\Delta\tau}+s^2e^{j\pi\alpha\Delta\tau}.
\end{equation}

\subsection{Fourth-Order SG-TED PMD Response}
\label{app:pmd_fourth_order}

From the first row of the PMD transfer matrix in Eq.~\eqref{eq:word_81}, the observed signal can be written in the time domain as

\begin{equation}
\label{eq:word_95}
x_{\mathrm{PMD}}(t)=\sum_k a_k h_x(t-kT_s-\tau)
+\sum_k b_k h_y(t-kT_s-\tau),
\end{equation}
where the PMD-induced effective pulse response is

\begin{equation}
\label{eq:word_96}
\begin{aligned}
h_x(t)&=c^2g\!\left(t-\frac{\Delta\tau}{2}\right)
+s^2g\!\left(t+\frac{\Delta\tau}{2}\right),\\
h_y(t)&=cse^{j\phi}
\left[g\!\left(t-\frac{\Delta\tau}{2}\right)
-g\!\left(t+\frac{\Delta\tau}{2}\right)\right].
\end{aligned}
\end{equation}
Define

\begin{equation}
\label{eq:word_97}
x_{\mathrm{PMD},+}=x_{\mathrm{PMD}}\!\left(t+\frac{u}{2}\right),\qquad
x_{\mathrm{PMD},-}=x_{\mathrm{PMD}}\!\left(t-\frac{u}{2}\right).
\end{equation}
For compactness, let

\begin{equation}
\label{eq:word_98}
\begin{aligned}
h_{x,k,\pm}&=h_x\!\left(t\pm\frac{u}{2}-kT_s-\tau\right),\\
h_{y,k,\pm}&=h_y\!\left(t\pm\frac{u}{2}-kT_s-\tau\right).
\end{aligned}
\end{equation}
Then,

\begin{equation}
\label{eq:word_99}
\begin{aligned}
x_{\mathrm{PMD},+}
&=\sum_k a_kh_{x,k,+}+\sum_k b_kh_{y,k,+},\\
x_{\mathrm{PMD},-}
&=\sum_k a_kh_{x,k,-}+\sum_k b_kh_{y,k,-}.
\end{aligned}
\end{equation}
The SG-TED operates on the power process. Hence, the relevant fourth-order quantity is

\begin{equation}
\label{eq:word_100}
\begin{aligned}
&\mathbb{E}\!\left[|x_{\mathrm{PMD},+}|^2|x_{\mathrm{PMD},-}|^2\right]\\
&=m_{\mathrm{PMD}}\!\left(t+\frac{u}{2}\right)
m_{\mathrm{PMD}}\!\left(t-\frac{u}{2}\right)\\
&\quad+\left|w_{x,\mathrm{PMD}}(t,u)\right|^2
+\kappa_a d_{\mathrm{PMD}}(t,u).
\end{aligned}
\end{equation}
The PMD-distorted second-order field-correlation kernel is

\begin{equation}
\label{eq:word_101}
w_{x,\mathrm{PMD}}(t,u)
=\sigma_a^2\sum_k
\left[
h_{x,k,+}h_{x,k,-}^*
+h_{y,k,+}h_{y,k,-}^*
\right],
\end{equation}
And the corresponding fourth-order pulse overlap kernel is

\begin{equation}
\label{eq:word_102}
d_{\mathrm{PMD}}(t,u)=\sum_k
\left[
|h_{x,k,+}|^2|h_{x,k,-}|^2
+|h_{y,k,+}|^2|h_{y,k,-}|^2
\right].
\end{equation}

Since $h_y(t)$ contains the relative phase $\phi$ only as a common factor $e^{j\phi}$, this phase cancels in $h_{y,k,+}h_{y,k,-}^{*}$ and does not appear in $|h_{y,k,\pm}|^2$. Therefore, $w_{x,\mathrm{PMD}}(t,u)$, $d_{\mathrm{PMD}}(t,u)$, and the resulting SG-TED CT strength are independent of $\phi$ under the assumed independent proper-complex tributaries.

For the SG-TED, the relevant lag is $u_0=T_s/2$. As in the no-PMD case, the average-power product term has no symbol-rate Fourier component at $u=u_0$. Therefore, the symbol-rate power process CAF is

\begin{equation}
\label{eq:word_103}
R_{p,\mathrm{PMD}}^{1/T_s}\!\left(\frac{T_s}{2}\right)
=W_{\mathrm{PMD}}^{1/T_s}\!\left(\frac{T_s}{2}\right)
+\kappa_aD_{\mathrm{PMD}}^{1/T_s}\!\left(\frac{T_s}{2}\right).
\end{equation}
Substituting this result into the SG-TED relation

\begin{equation}
\label{eq:word_104}
C_{\mathrm{SG}}^{\mathrm{PMD}}
=-2jR_{p,\mathrm{PMD}}^{1/T_s}\!\left(\frac{T_s}{2}\right),
\end{equation}
we obtain the CT power under first-order PMD as

\begin{equation}
\label{eq:word_105}
\begin{aligned}
G_{\mathrm{SG}}^{\mathrm{PMD}}
&=4\left|
W_{\mathrm{PMD}}^{1/T_s}\!\left(\frac{T_s}{2}\right)
+\kappa_aD_{\mathrm{PMD}}^{1/T_s}\!\left(\frac{T_s}{2}\right)
\right|^2\\
&=4|C_{\mathrm{PMD}}+C_{\kappa,\mathrm{PMD}}|^2.
\end{aligned}
\end{equation}

\end{document}